\documentclass[a4paper,fleqn,usenatbib,useAMS]{mnras}

\usepackage{newtxtext,newtxmath}

\usepackage[T1]{fontenc}
\usepackage{ae,aecompl}

\usepackage{graphicx}	
\usepackage{amsmath}	
\usepackage{amssymb}	
\usepackage{multicol}        
\usepackage{bm}		
\usepackage{pdflscape}	

\title[The Environmental Dependencies]{XMM-Newton view of X-ray overdensities from nearby
galaxy clusters: the environmental dependencies}

\author[Caglar \& Hudaverdi]{
Turgay Caglar,$^{1}$\thanks{E-mail: turgay.caglar@std.yildiz.edu.tr}
Murat Hudaverdi,$^{2,1}$
\\
$^{1}$Department of Physics, Yildiz Technical University, Davutpasa Campus, 34220, Istanbul, Turkey\\
$^{2}$AUM, College of Engineering and Technology, Department of Science, Dasman 15453, Kuwait
}

\date{Accepted 2017 July 17. Received 2017 June 19; in original form 2016 October 29}

\pubyear{2017}
   \volume{471}

\begin{document}
\label{firstpage}
\pagerange{4990--5007}
\maketitle

\begin{abstract}

In this work, we studied ten nearby (z$\leq$0.038) galaxy clusters to understand possible interactions between hot plasma and member galaxies. A multi-band source detection was applied to detect point-like structures within the intra-cluster medium. We examined spectral properties of a total of 391 X-ray point sources within cluster's potential well. Log \textit{N} - Log \textit{S} was studied in the energy range of 2-10 keV to measure X-ray overdensities. Optical overdensities were also calculated to solve suppression/triggering phenomena for nearby galaxy clusters. Both X-ray to optical flux/luminosity properties, (\textit{X/O}, \textit{L$_{X}$/L$_{B}$}, \textit{L$_{X}$/L$_{K}$}), were investigated for optically identified member galaxies. X-ray luminosity values of our point sources are found to be faint (40.08 $\leq$ log(L$_{X}$) $\leq$ 42.39 erg s$^{-1}$). The luminosity range of point sources reveals possible contributions to X-ray emission from LLAGNs, X-ray Binaries and star formation. We estimated $\sim$ 2 times higher X-ray overdensities from galaxies within galaxy clusters compared to fields. Our results demonstrate that optical overdensities are much higher than X-ray overdensities at the cluster's centre, whereas X-ray overdensities increase through the outskirts of clusters. We conclude that high pressure from the cluster's centre affects the balance of galaxies and they lose a significant amount of their fuels; as a result, clustering process quenches X-ray emission of the member galaxies. We also find evidence that the existence of X-ray bright sources within cluster environment can be explained by two main phenomena: contributions from off-nuclear sources and/or AGN triggering caused by galaxy interactions rather than AGN fuelling.
\end{abstract}



\section{INTRODUCTION}
Clusters of galaxies are formed by gravitational infalling of smaller structures, and thus they are observed to be in high density regions of the Universe. 
Their deep potential well retains hot gas and individual galaxies in the vicinity.
The morphology and star formation rate (SFR) of such infalling galaxies change as a result of their interaction with the intra-cluster medium (ICM). 
Possible interactions and collisions between member galaxies are very likely probable. All these complexities can be effective on the galaxy evolution within galaxy clusters. The advents of the technology in space science allow us to study the evolution of galaxies in these dense and complex regions.  

Several studies at different redshifts report X-ray source overdensities from galaxy clusters
 \citep[e.g.,][]{Cappi,D'Elia,Hudaverdi,Gilmour,Koul,Ehlert}. 
The comparison between clustered and non-clustered fields has been very successful to explain the nature of X-ray point sources. 
However, it is still unclear whether cluster environments suppress or enhance X-ray active galactic nuclei (AGN) activity.
\citet{Koul} attempted to answer this issue by comparing X-ray and optical overdensities from 16 rich Abell clusters 
and reported a strong suppression within the dense ($<$ 1 Mpc) cluster environment.  
\citet{Khabiboulline} studied low redshift clusters ($z <$ 0.2) and showed that AGN activity is suppressed in the rich cluster centre.  
\citet{Haines} also confirmed a similar result for massive clusters. 
On the other hand, radially moving outward to the cluster outskirts, \citet{RE} showed an enhancement of X-ray AGN activity for 51 clusters within 3.5 Mpc.
This result is also confirmed for distant clusters ($z >$ 1) by further investigations \citep[]{Fassbender, Koul1, Koul2, Alberts}.
A recent study found evidence that AGN emission is found to be strongly related to the richness class of the host cluster. 
It is understood that rich clusters suppress X-ray AGN activity \citep[e.g.,][]{Koul,Ehlert,Haines,Koul1}.
On the similar topic, \citet[][]{Bufanda} did not, however, find any clear correlation between AGN fraction and 
cluster richness based on a study of 432 galaxy clusters' data in the redshift range 0.10 $ < z <$ 0.95. Therefore, The role of environment in the frequency of AGN is still an open question. A number of studies demonstrate an increased nuclear activity of the galaxies in the rich cluster environment. \citet{Martini2006} verified the existence of large low-luminous active galactic nuclei (LLAGN) populations and reported the fraction as $\sim$5$\%$ in the nearby galaxy clusters. Furthermore, \citet{Melnyk} reported that 60\% of X-ray selected AGNs are found to be in dense environments and thus likely to reside in clusters of galaxies. \cite{Ellison} reported substantial evidence of increased AGN activity due to close encounters of galaxies in the gravitational potential well of the host cluster. \citet{Haggard} estimated approximately equal optical AGN fraction from clusters relative to the fields. \citet[][]{Ehlert,Ehlert2} found X-ray AGN fraction of 42 massive cluster centres to be three times lower than the fields. Traditional optical studies reveal a lower optical AGN fraction from clusters; the fraction for cluster and non-cluster fields is found to be $\sim$1$\%$ and  $\sim$5$\%$, respectively \citep[e.g.,][]{Dressler1999}. Recent studies also confirm that optically bright AGNs are rare in cluster environments \citep[e.g.,][]{Kauffmann,PoBi}.  

The main astrophysical objects responsible for X-ray emission are diffuse hot gas, X-ray Binaries (XRBs) and accreting supermassive black holes (SMBHs); therefore X-ray emission mechanisms are highly related to the dynamic events occurring within the galaxy. In the case of absence of very luminous X-ray sources, galaxy X-ray emission fainter than Lx $<$ 10$^{42}$ erg s$^{-1}$ can be produced either from star formation activities or LLAGNs. Recent studies imply that X-ray emission from the majority of LLAGNs can be related to off-nuclear sources or diffuse emission rather than central nuclear emission \citep[e.g.,][]{Ho,Ranalli,Ranalli2012}. On the other hand, \citet{Gisler} provided a correlation between star-formation and dense environments: galaxies have low star formation rates in crowd regions. Recent studies also confirmed this relationship \citep[e.g.,][]{Kauffmann,Schaefer}. Observed low star formation rates from corresponding galaxies are highly relative to the distance from central regions of clusters and associated with environmental suppression \citep[e.g.,][]{Oemler,Balogh,Wetzel}. To understand properties of the star forming galaxies, some indicators have been derived from multi-wavelength surveys \citep{Ranalli,Mineo}. 

In this study, we aim to understand the contribution of environment to the galaxy evolution and interaction between ICM and member galaxies. We also intend to measure X-ray and optical density of selected galaxy clusters. There is a conflict whether galaxy clustering process suppresses or enhances galaxy X-ray activity. We attempt to solve this conflict in nearby clusters by searching for X-ray and optical overdensities relative to fields. We selected a sample of several nearby galaxy clusters ($\leq$ 171 Mpc). However, in bright galaxy clusters, faint X-ray point sources cannot be detected in very bright ICMs. In that case, X-ray source number densities can be decreased. To overcome this effect, we concentrated on faint galaxy clusters with unextended ICM emission (r$_{c}$ $<$ 170 kpc). Our paper is organised as follows: Section 2 reviews observational samples and the data reduction process. Section 3 describes how we performed X-ray and optical analysis. Section 4 describes our measurement method for X-ray and optical overdensities. In section 5, we discuss our results in two different topics: contribution to X-ray emission from LLAGNs and star formation. Section 6 concentrate on the nature of X-ray and optical emission from individual galaxies. Finally, in section 7, we present our conclusions. We adopt WMAP standard cosmological parameters H$_{0}$ = 70 km s$^{-1}$ Mpc$^{-1}$, $\Omega$$_{M}$ = 0.27 and $\Omega$$_{\Lambda}$ = 0.73 in a flat universe. \\

\section{OBSERVATION AND DATA REDUCTION}
We used archival data of the \textit{XMM-Newton} in our analysis, and all observational data were gathered from \textit{XMM-Newton} Science Archive (XSA). In our survey, we concentrated on selecting the \textit{XMM-Newton} observational data that were taken in full frame mode for MOS and extended full frame mode for pn. X-ray observation data logs are listed in Table $\ref{t1}$. 

The \textit{XMM-Newton} data were processed by using \textit{heasoft 6.19} and \textit{XMMSAS 15.0.0} current calibration files (ccf) and summarised observation data files (odf) were generated by using \textit{cifbuild-4.8} and \textit{odfingest-3.30} respectively. We generated event files using \textit{epchain-8.75.0} and \textit{emchain-11.19} tasks from the observation data file. Rate filter is applied to the event file to clear flaring particle background. 

\begin{table}
  \resizebox{0.46\textwidth}{!}{\begin{minipage}{0.49\textwidth}
        \caption{XMM-Newton observation logs of our sample of clusters.}
  \begin{tabular}{cccc}
\hline
\hline
 Obj. Name & Obs. ID & Obs. Date & Exp. Time (ks)  \\
& & & M1 M2 PN \\
 \hline
\vspace{0,5mm}
Abell 3581 & 0504780301 & 01/08/2007 & 117  117 113  \\
\vspace{0,5mm}
Abell 1367 & 0061740101 & 26/05/2001 & 33 33 28  \\
\vspace{0,5mm}
Abell 1314 & 0149900201 & 24/11/2003 & 18 18 17  \\
\vspace{0,5mm}
Abell 400 & 0404010101 & 06/08/2006 & 39 39 33  \\
\vspace{0,5mm}
Abell 1836 & 0610980201 & 17/01/2010 & 37 37 35  \\
\vspace{0,5mm}
Abell 2063 & 0550360101 & 23/07/2008 & 28 28 24  \\
\vspace{0,5mm}
Abell 2877 & 0204540201 & 23/11/2004 & 22 22 20  \\
\vspace{0,5mm}
Abell S137 & 0744100101 & 16/05/2014 & 27 27 32  \\
\vspace{0,5mm}
Abell S758 & 0603751001 & 21/02/2010 & 64 64 60  \\
\vspace{0,5mm}
RXCJ2315.7-0222  & 0501110101 & 22/11/2007 & 44 44 40  \\
\vspace{0,5mm}
Deep 1334+37 & 0109661001 & 23/06/2001 & 86 86 86  \\
\vspace{0,5mm}
Groth-Westphal & 0127921001 & 21/07/2000 & 56 56 52  \\
\vspace{0,5mm}
Hubble Deep N & 0111550301 & 27/05/2001 & 46 46 45  \\
\hline
\label{t1}
\end{tabular}
\end{minipage}}
\end{table}

\section{ANALYSIS}
\subsection{Spatial and Spectral Analysis}
We applied SAS source detection algorithms to the data. Source detection was performed by using SAS task, namely \textit{edetect\_chain-3.14.1}. We used five different images in the super soft band (0.2-0.5 keV), in the soft band (0.5-1.0 keV), in the medium band (1.0-2.0 keV), in the hard band (2.0-4.5 keV), and in the super hard band (4.5-12.0 keV) for source detection. Source detections were accepted with likelihood values above 10 (about 4 $\sigma$) and inside an off-axis angle of 12.5$\arcmin$. Detection routine was applied for both mos and pn cameras, and the final list was combined with sas task '\textit{srcmatch-3.18.1}'. 
After detecting point-like sources, spectral and background files were produced by using sas task \textit{evselect-3.62}. The background spectrum was extracted from an annulus surrounding the circular source region. Area of spectral files was calculated by using \textit{backscale-1.4.2}. The Redistribution Matrix Files and Ancillary Response Files were produced by using SAS tasks \textit{rmfgen-2.2.1} and \textit{arfgen-1.92.0} respectively. The spectra of a majority of the point sources were modelled with a single absorbed power-law. However, the spectra of several sources contain thermal emission lines that cannot be fit well by using a single power law. In that situation, we added a thermal component (APEC) to improve fitting. The average intra-galactic abundance value was fixed at 0.3 solar value in our analysis \citep[][]{Getman}. 

\subsection{Sensitivity of the Survey} \label{s3.3}
The sky coverage represents the survey area of the observed source and decreases with flux due to instrumental effects. Therefore, limiting flux of our survey needs to be calculated very carefully. There are a few factors that affect limiting flux, such as point spread function, vignetting, exposure time, and detector sensitivity. We calculated the sensitivity of our cameras by using sas task \textit{esensmap-3.12.1}. The energy conversion factors of our samples (ECF) were calculated from rate/flux by considering hydrogen column density, photon index, and filter type of operating camera. ECF values of our samples were calculated with XSPEC model (wabs*power) with fixed photon index of 1.7 and fixed total galactic hydrogen column density value. Resulting ECF values and limit flux of our samples are presented in Table $\ref{t2}$.
Since galaxy clusters emit centrally concentrated very diffuse X-ray emission, the detection of the faint sources buried inside the ICM is not possible. To overcome this problem, we did not take into account the central region of our sample of clusters ($\sim$ 3 $\times$ r$_{c}$) in our analysis. Central regions of each cluster (95 $<$ r$_{c}$ $<$ 145 h$^{-1}_{70}$ kpc) were calculated from King's Profile \citep{King}. Due to these selection techniques, X-ray sources fainter than 1 $\times$ $10^{-14}$ erg cm$^{-2}$ s$^{-1}$ were not taken into consideration in our analysis, and we also didn't present their properties in the appendix.

\begin{table}
  \resizebox{0.46\textwidth}{!}{\begin{minipage}{0.49\textwidth}
        \caption{Detection sensitivity survey: I) Name of the galaxy cluster II) Energy conversion factor III) A total number of detected sources IV) Final source number V) Flux limit of the corresponding cluster.}
  \begin{tabular}{ccccc}
\hline
\hline
I & II & III & IV & V \\
Cluster & ECF & N$_{T}$ & N$_{F}$  & Limit Flux \\
& cts cm$^{2}$ erg$^{-1}$ & && erg cm$^{-2}$ s$^{-1}$ \\
 \hline
Abell 1367 & 4.72$\times$$10^{11}$ & 71 & 33 & 6.76$\times$$10^{-15}$ \\

Abell 3581 & 4.91$\times$$10^{11}$ & 105 & 47 & 4.07$\times$$10^{-15}$ \\ 

Abell 400 & 3.81$\times$$10^{11}$ & 62 & 33 & 6.17$\times$$10^{-15}$ \\

Abell 2877 & 4.78$\times$$10^{11}$ & 69 & 35 & 5.50$\times$$10^{-15}$ \\

Abell S137 & 5.07$\times$$10^{11}$ & 91 & 54 & 3.80$\times$$10^{-15}$ \\

Abell 1314 & 4.46$\times$$10^{11}$ & 111 & 39 & 5.25$\times$$10^{-15}$ \\

Abell 2063 & 5.21$\times$$10^{11}$ & 34 & 25 & 7.59$\times$$10^{-15}$ \\

Abell 1836 & 4.17$\times$$10^{11}$ &116 & 32 & 6.31$\times$$10^{-15}$ \\

Abell S758 & 4.58$\times$$10^{11}$ & 130 & 59 & 2.88$\times$$10^{-15}$ \\

RXCJ2315.7-0222 & 4.25$\times$$10^{11}$ & 85 & 34 & 4.17$\times$$10^{-15}$ \\
\hline
\label{t2}
\end{tabular}
\end{minipage}}
\end{table}

\subsection{Optical Data} \label{s3-3}
Even though a small number of red spirals and blue ellipticals are reported in the literature \citep[e.g.,][]{vandenbergh,Masters}, spiral galaxies are typically found in blue clouds, while ellipticals are usually on the red sequence \citep[e.g.,][]{Tully1982,Kauffmann2003,Tojeiro}. The fraction of early-type galaxies with respect to the whole galaxy population is significantly higher in clusters than in the field \citep[e.g.,][]{Oemler, Dressler, Dressler1999, Kauffmann}, whereas the number of the blue-type galaxies increases towards the outskirts of the clusters \citep[e.g.,][]{Butc-Oem1,Pimbblet}. In this section, we aim to identify the colour of galaxies (blue/red) within cluster fields. 

\begin{figure}
\begin{center}
\includegraphics*[width=8.4cm,angle=0]{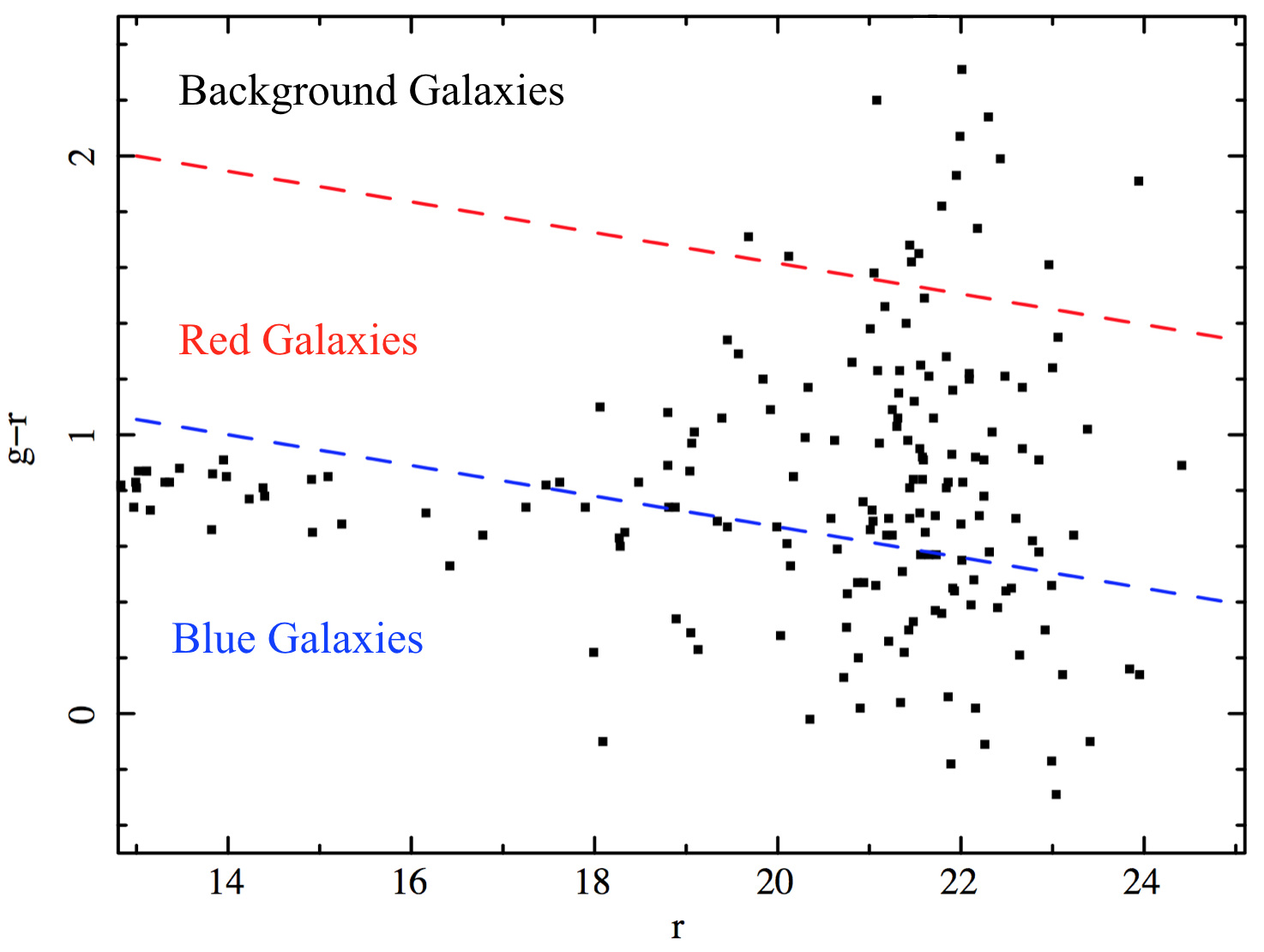}
\includegraphics*[width=8.4cm,angle=0]{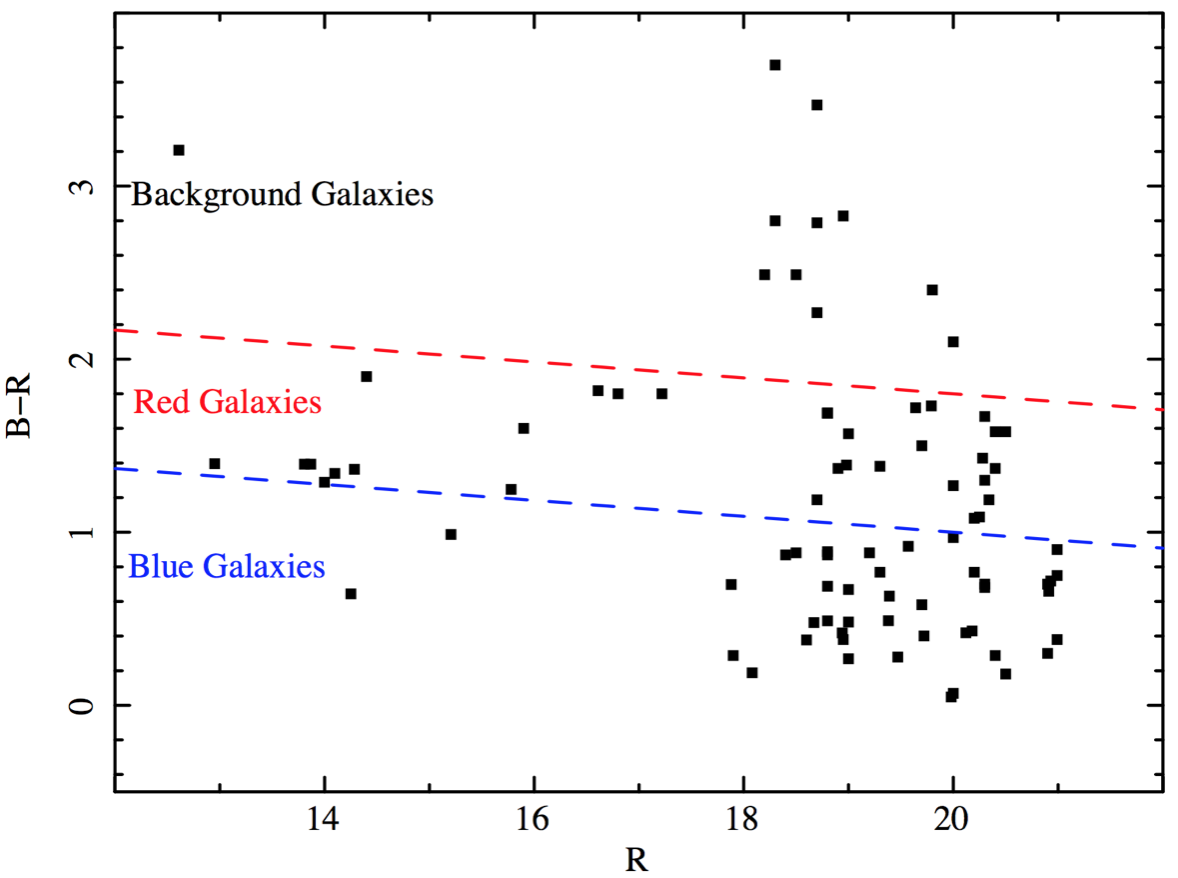}
\end{center}
\caption{Colour-magnitude diagram of galaxies within our sample of clusters.}
\label{1}
\end{figure}

Optical counterparts of X-ray sources are identified from SDSS archive. However, we note that not all X-ray sources have optical counterparts. Also, X-ray centroid of galaxies does not always coincide with optical centroids. It is well known that major events, such as clusters mergers or tidal interactions, cause offset between X-ray and optical centre \citep[e.g.,][]{Peres}. \citet[e.g.,][]{Loaring} demonstrated the existence of a trend between flux and positional error for \textit{XMM-Newton} point sources and reports maximum positional error of \textit{XMM-Newton} for faint sources as $<$ 10$\arcsec$ within off-axis angle 9$\arcmin$; moreover, the positional error of sources becomes higher at the off-axis angle > 9$\arcmin$. Owing to these assumptions, we considered optical counterpart of X-ray sources within $<$ 6$\arcsec$ ($<$ 4.7 kpc) radius. Then, likelihood ratio for each candidate is computed by using cross-correlation method described by \citet[e.g.][and references therein]{Pineau}. Finally, sources falling outside the likelihood ratio < \%50 are assumed as background sources. We also mention that similar methods were applied to different surveys \citep[e.g.,][]{Brusa,Flesch,Pineau,LaMassa}. 
We exhibit galaxy r band magnitudes as a function of g-r and b-r in Fig. $\ref{1}$. Magnitude values were taken from SDSS archive for the following galaxy clusters: A400, A1314, A1367, A1836, A2063, and RXCJ2315.7-0222. However, there are no SDSS observations for the rest of the galaxy clusters. To study them, we used three different catalogues to gather b and r band magnitudes of galaxies \citep[][]{Flesch,Zach2005,Zach2013}. We also note that the K and extinction correction are applied to all magnitude values unless they are noted as extinction corrected. Dashed lines represent the limit value to separate blue and red galaxies \citep[][]{Omar,Lagan}. We identified a control zone using g-r $\pm{0.4}$ limit for background galaxies. We assume that the galaxies falling outside of upper limit of the red sequence are unrelated to galaxy clusters.	 

\section{OVERDENSITY MEASUREMENTS} \label{s4}
\begin{figure}
\begin{center}
\includegraphics*[width=8.13cm,angle=0]{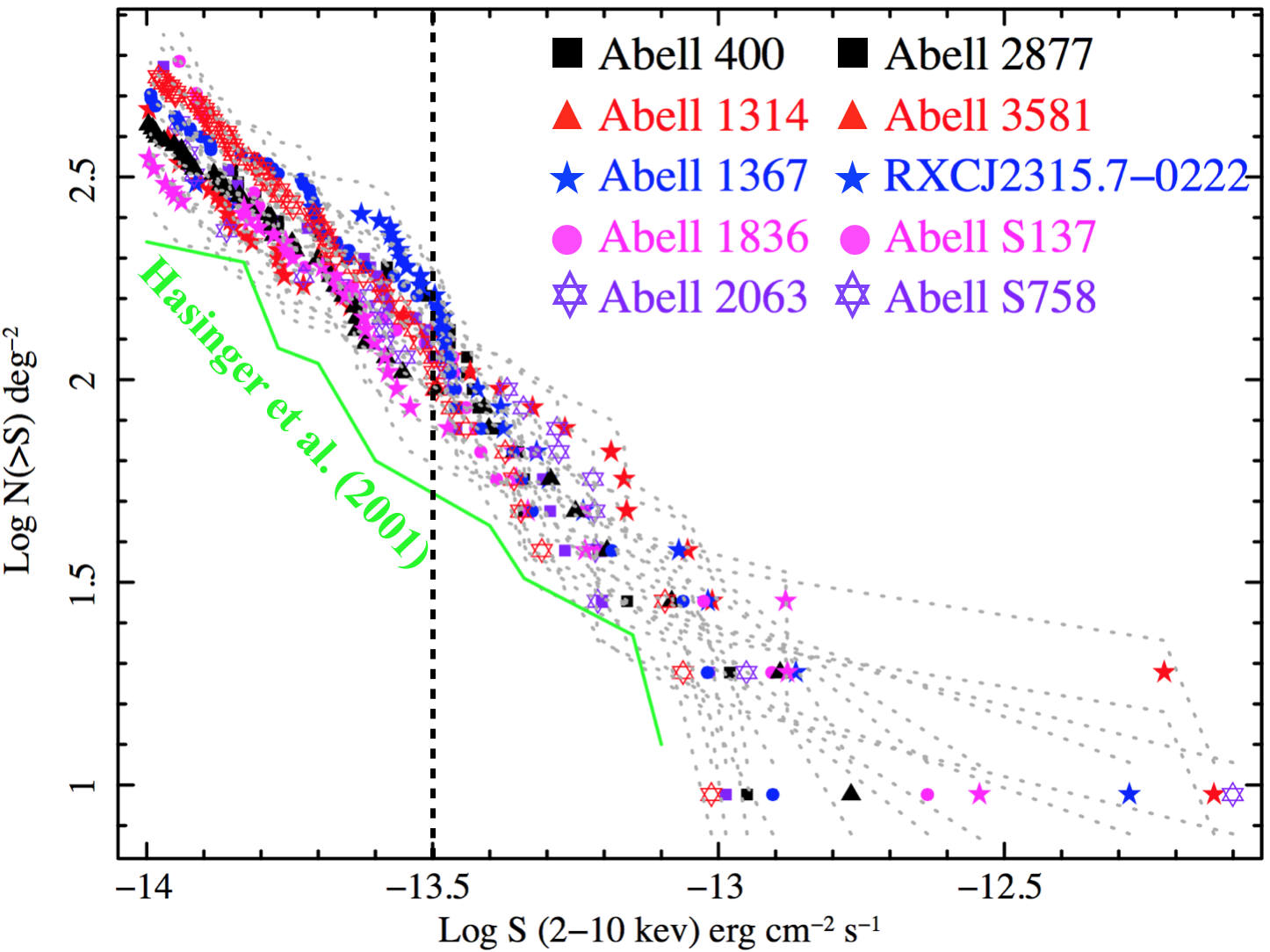}
\includegraphics*[width=8.1cm,angle=0]{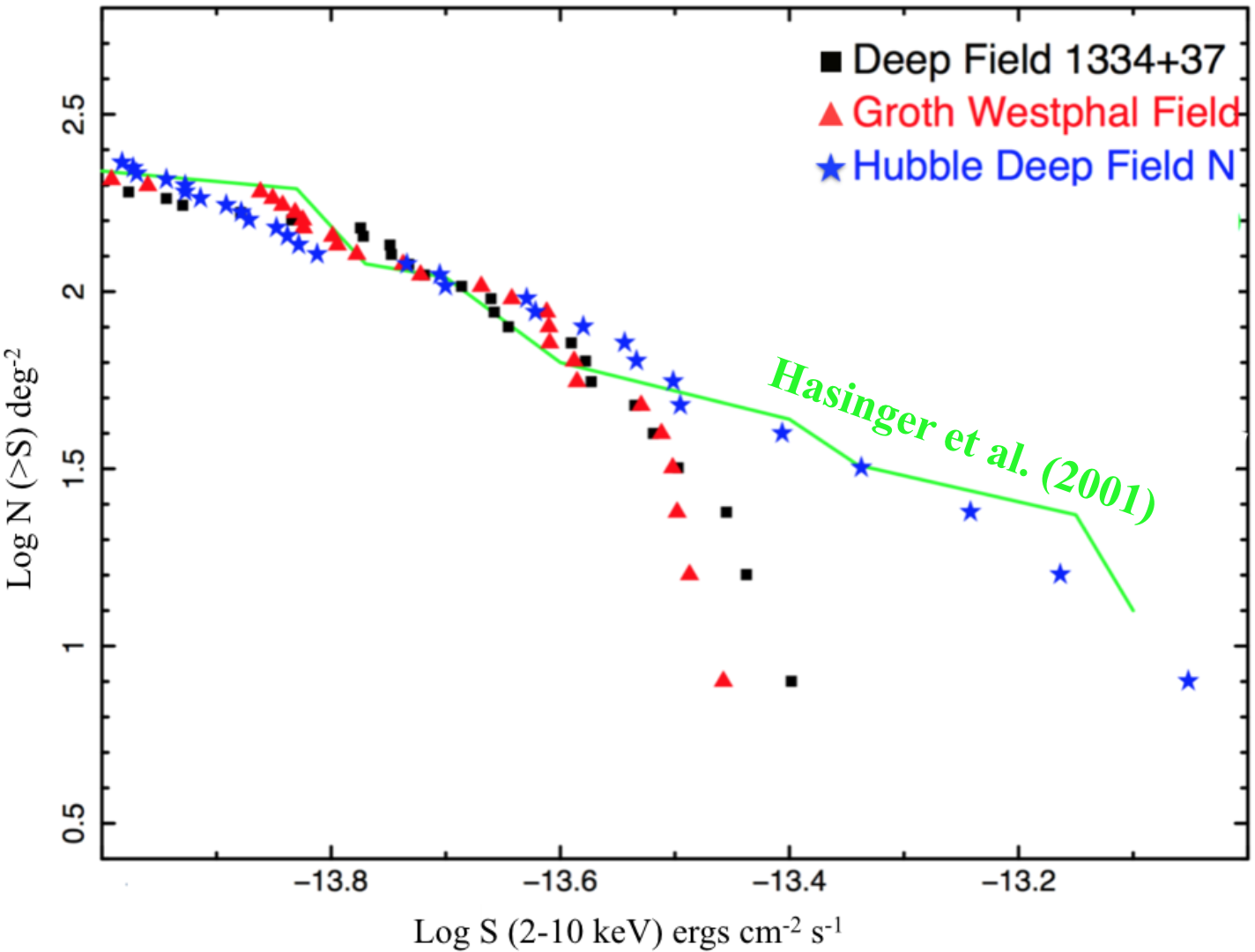}
\end{center}
\caption{Log \textit{N} - Log \textit{S} calculated in the 2-10 keV band for our sample of clusters (\textbf{Top}) and our sample of fields (\textbf{Bottom}). The grey dotted lines represent a $1\sigma$ statistical error. The black dashed lines demonstrate log(f) = -13.5 erg cm$^{-2}$ s$^{-1}$ value for visual aid, which we used in our X-ray overdensity measurements. The green lines represent Lockman Hole results, which is studied by \citet{Hasinger}.}
\label{2}
\end{figure}

The number of sources per unit sky area with the flux higher than S, N ($>S$), is defined as ;

\begin{equation}
N(>S) = \sum_{i = 1}^{n} \frac{1}{\Omega_{i}} deg^{-2}
\end{equation}

where n is number of detected sources, ${\Omega_{i}}$ is sky coverage for the flux of the i-th source. Fig. $\ref{2}$ shows log \textit{N} - log \textit{S} for our samples and their comparison with Lockman hole result, which was calculated by \citet{Hasinger}. Several studies show that cosmic variance within 2-10 keV energy range is less than 15\% \citep[e.g.,][]{Cappelluti,Dai}; hence, we selected 2-10 keV flux values for our log \textit{N} - log \textit{S} measurements. In our survey, sources brighter than log(f$_{\rm 2-10 keV}$) = -13.5 erg cm$^{-2}$ s$^{-1}$ are not affected by decreasing of sky coverage, with this; we calculate X-ray overdensities at this particular flux value. At log(f$_{\rm 2-10 keV}$) = -13.5 erg cm$^{-2}$ s$^{-1}$, \citet{Hasinger} estimated 52$\pm{7}$ sources per degree square for the Lockman Hole Field.  We calculated 53$\pm{8}$ sources per degree square for Hubble Deep Field North at this flux value. X-ray source overdensities have been computed using the equation 1+$\delta_{x}$ = $N_{x}$/$N_{e}$ \citep{Koul}, where $N_{x}$ is the number of X-ray sources brighter than log \textit{S} (-13.5 erg cm$^{-2}$ s$^{-1}$) and $N_{e}$ is expected X-ray source numbers from non-clustered fields. 
Optical overdensities were also calculated from the following equation 1+$\delta_{o}$ = $N_{o}$/$N$. In this formula, $N_{o}$ is the number of objects with the characteristic magnitude of the selected galaxy cluster within the field of view and N is the total number of objects from non-clustered fields with the same characteristic magnitude. Due to minimisation of projection effects, all galaxies around the cluster centre were extracted by using the method explained by \citet[][]{Koul} (see section $\ref{s3.3}$). Eventually, we calculated optical galaxy overdensities by using characteristic magnitude approximation with optical data. The characteristic magnitude of our clusters within range of m$^{*}$$_{r}$ $\pm{2.0}$ was estimated by using the following equation m$^{*}$ = M$^{*}$ + 5log(d) + K(z) + 25 + A$_{v}$, where M$^{*}$ is fit parameter from Schechter Luminosity function for r band \citep[][]{Montero-Dorta},  A$_{v}$ is the galactic absorption, which is estimated from galactic absorption map \citep[][]{Schlafly}, and K(z) is the K-correction factor \citep[][]{Poggianti}. We also remind that classified stars and foreground/background galaxies were not taken into consideration in our analysis. Overdensity results of our samples are presented in Table $\ref{t3}$.

\begin{table}
  \resizebox{0.48\textwidth}{!}{\begin{minipage}{0.50\textwidth}
        \caption{Our sample of clusters: X-ray overdensities. $\star$: N($>$S) values at log(f$_{\rm 2-10 keV}$) = -13.5 erg cm$^{-2}$ s$^{-1}$.}
  \begin{tabular}{ccccc}
  
\hline
\hline
Cluster & Redshift &  $\delta_{X}$  & m$_{r}$$^{*}$ & $\delta_{o}$   \\
&  & & mag &  \\
 \hline
\vspace{0,5mm}
A400 & 0.024  & 1.67$\pm{0.75}$ & 14.80 & 3.50$\pm{0.87}$ \\
\vspace{0,5mm}
A1314 & 0.034  & 1.16$\pm{0.53}$ & 15.24 & 5.33$\pm{1.02}$ \\
\vspace{0,5mm}
A1367 & 0.022  & 1.67$\pm{0.75}$ & 14.45 & 3.00$\pm{0.82}$ \\
\vspace{0,5mm}
A1836 & 0.036  & 1.00$\pm{0.45}$ & 15.54 & 3.10$\pm{0.64}$ \\
\vspace{0,5mm}
A2063 & 0.035  & 1.00$\pm{0.45}$ & 15.34 & 4.38$\pm{0.82}$ \\
\vspace{0,5mm}
RXCJ2315.7-0222 & 0.027 & 0.33$\pm{0.15}$ & 14.61 & 2.00$\pm{0.71}$ \\
\vspace{0,5mm}
A2877 & 0.025 & 1.16$\pm{0.53}$ & 14.39 & none \\
\vspace{0,5mm}
A3581 & 0.023 &  0.67$\pm{0.30}$ & 14.54 & none \\ 
\vspace{0,5mm}
AS137 & 0.026 & 1.67$\pm{0.75}$ & 14.63 & none \\
\vspace{0,5mm}
AS758 & 0.038 & 1.16$\pm{0.53}$ & 15.64 & none \\
\hline
\label{t3}
\end{tabular}
\end{minipage}}
\end{table}

\vspace{2.5mm}

\section{DISCUSSION}
\begin{figure*}
\begin{center}
\includegraphics*[width=8.29cm,angle=0]{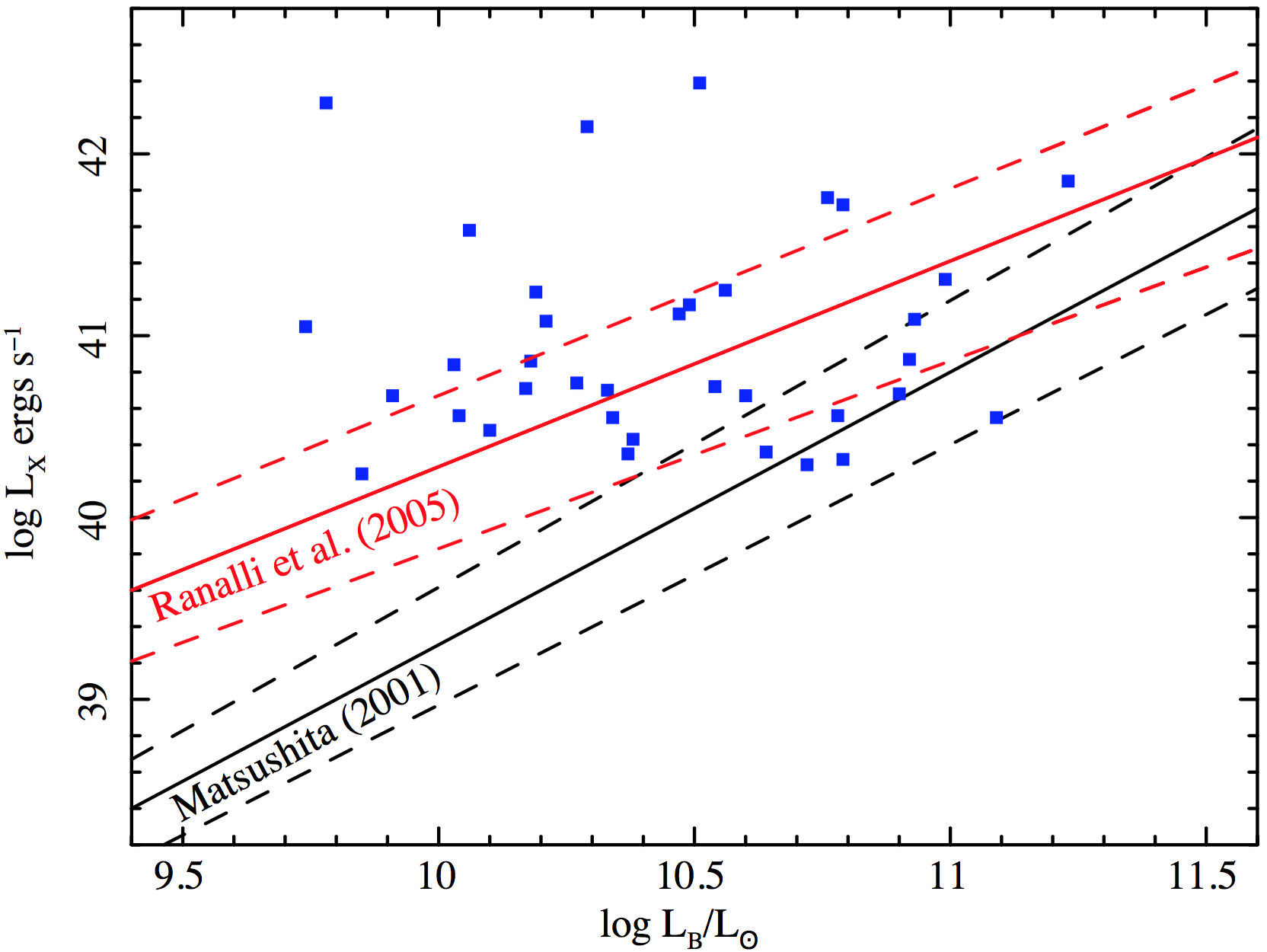}
\includegraphics*[width=8.31cm,angle=0]{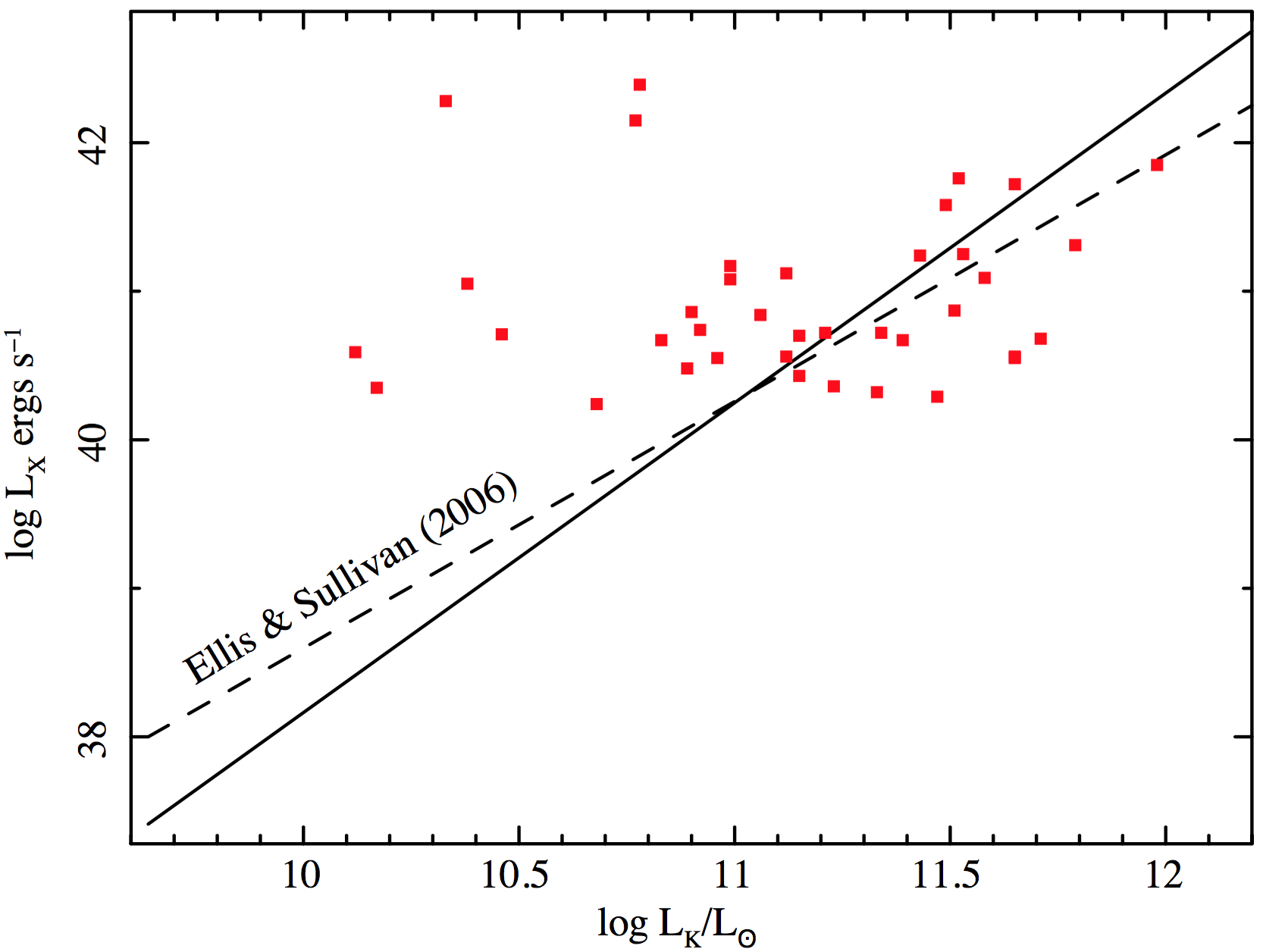}
\caption{\textbf{Top:} L$_{X}$ as a function of L$_{B}$ for our galaxies within the cluster environment. The black solid line represents the expected distribution of early-type galaxies reported by \citet{Matsushita}, while the black dashed lines mark the $\pm$10\% uncertainties on the relation. The red solid line represents the expected distribution of late type galaxies reported by \citet{Ranalli2005}, while the red dashed lines mark the $\pm$8\% uncertainties on the relation. \textbf{Bottom:} L$_{X}$ as a function of L$_{K}$ for our galaxies within the cluster environment, compared to results reported by \citet{ES}. The black solid line shows the best-fit to their early type galaxies sample, while the black dotted line represents the expected distribution of early-type galaxies in non-clustered fields.}
\label{3}
\end{center}
\end{figure*}

We performed data analysis for \textit{XMM-Newton} observations of a sample of clusters and fields. Log \textit{N} - Log \textit{S} was studied at limiting flux value of 1 $\times$ $10^{-14}$ erg cm$^{-2}$ s$^{-1}$ and, we found $\sim$ 2 times higher X-ray source density from our clusters compared to the values calculated in the Hubble Deep Field North and those estimated in the Lockman hole field studied by \citet{Hasinger} (see Fig. \ref{2}). Even in the worse case scenario according to the error limits, at least \% 35 of our point sources are cluster members. Due to minimise the influence of ICM, we did not take into account the central regions of our clusters in our analysis. Also, we studied three different fields to enlarge our knowledge about non-clustered fields. The number counts, which were calculated in the Lockman Hole and our field samples are consistent with each other, and we confirm lower X-ray source densities in non-clustered fields than in galaxy clusters (see Fig. $\ref{2}$). Encouraged by this result, we calculated X-ray to optical flux ratio to understand the variety of X-ray sources detected in clusters. R-band magnitudes were compared to hard X-ray flux values and \textit{X/O} were calculated by using the equation \textit{X/O} = log(f$_{X}$) + C + m$_{opt}$ $\times$ 0.4 \citep{Maccacaro}. R-band magnitudes are taken from SDSS, NOMAD and MORX catalogue, and we applied extinction correction by using extinction maps from \citet{Schlafly}. Comparison between r-band magnitudes and X-ray fluxes is an advantageous method to address  the condition of nuclear activity/inactivity of galaxies. Whereas AGNs tend to have \textit{X/O}$>$-1 \citep[e.g.,][]{Fiore}, normal galaxies have \textit{X/O} $<$-2 \citep[e.g.,][]{Xue}. Besides, galaxies with -2 $<$ \textit{X/O} $<$ -1 value can either be LLAGNs or star-forming galaxies \citep{Park}. However, we cannot completely explain the type of the source due to the X-ray versus optical flux ratio. Therefore, we calculated X-ray to optical luminosity ratio for 40 member galaxies, and the results are given in Table $\ref{t5}$. \citet{Matsushita} studied early type galaxies and found the expected L$_{X}$/L$_{B}$ distribution of normal early type galaxies. \citet{Ranalli2005} also reported the expected L$_{X}$/L$_{B}$ distribution of late-type galaxies. We compared hard band X-ray luminosities (2 -10 keV) to blue optical luminosities to understand the behaviour of our cluster member galaxies. Absolute magnitudes were computed using the equation: M$_{opt}$ = m$_{opt}$ + 5 - 5log(d), where d is the distance in parsec unit, m$_{opt}$ is apparent magnitude value. Optical luminosities in solar units were calculated by using the equation log(L$_{opt}$/L$_{\odot}$) = -0.4 $\times$ (M$_{opt}$ - C), where C is  absolute magnitude of the sun in the related band. The majority of our sources has significantly higher L$_{X}$/L$_{B}$ than early type galaxies. Furthermore, $\sim$ \%50 of our member galaxies follow expected L$_{X}$/L$_{B}$ distribution of late type galaxies. Based on our L$_{X}$/L$_{K}$ results, we found that majority of the member galaxies is brighter in the X-rays than they are in the K band. The trends with L$_{B}$ and L$_{K}$ of L$_{X}$ plots are presented in Fig. $\ref{3}$. Moreover, we assumed our point sources  as likely cluster member and calculated luminosity values of our point sources by using cluster's redshifts. The luminosity range of our X-ray sources are found to be faint (40.08 $\leq$ Log(L$_{X}$) $\leq$ 42.39 erg s$^{-1}$). In this luminosity range, the X-ray emission can be produced by either LLAGNs, star formation and unresolved XRBs. We also note that the majority of X-ray sources of our survey is found to be normal or star-forming galaxies (log(L$_{X}$) < 41.00 erg s$^{-1}$) (see Tables $\ref{t7}$, $\ref{t8}$, $\ref{t9}$, $\ref{t10}$, $\ref{t11}$, and $\ref{t12}$). This result implies no central nuclear activity from these sources. Early studies of the local Universe demonstrate that XRB populations dominate X-ray emission from normal galaxies \citep[e.g.,][]{Muno}, which can be the main X-ray emission mechanism of normal galaxies in our survey. 

\subsection{The contribution from LLAGNs}
Based on our results, there is a possibility of AGN fuelling and quenching scenario in our clusters. When a galaxy falls into cluster environment under the influence of gravitational potential, the surrounding gas powers AGN \citep{Lietzen}, therefore, the source becomes brighter. Most of the galaxies host a black hole at the centre \citep{KR}, and possible fuelling from ICM activates inactive Black Holes \citep[e.g.,][]{AH}. Besides, close encounters and collisions of galaxies are highly probable in cluster environments, where close encounters possibly cause AGN triggering \citep[][]{Ellison}. Our results show the suppression of X-ray AGNs in the central regions of clusters (see Fig. \ref{4}, \ref{5}, \ref{6}). It appears that high pressurised winds from the cluster's centre affect the balance of galaxies within cluster environment and cause them to lose significant amounts of their fuel. This mechanism also explains the absence of very luminous galaxies at L$_{X}$ $\geq$ 10$^{42}$ erg s$^{-1}$ in nearby clusters. In this case, LLAGNs in nearby cluster environments can be related to close encounters of galaxies rather than AGN fuelling. 

\subsection{Star formation}
 A considerable number of recent studies reports that star formation rate increases through cluster outskirts, however star formation rates of galaxies are still lower than the field even at the viral radius of clusters of galaxies \citep[e.g.,][]{Balogh2,Lewis,Muzzin,Wagner}. On the other hand, it is well known that old red galaxies dominate cluster centres \citep[e.g.,][]{Dressler}; however, blue galaxies are also commonly detected in clusters. Several studies indicate that there is a significant relation between galaxy colour and star formation \citep[e.g.,][]{Tojeiro}, whereas blue galaxies with high SFR are bright in X-ray \citep[e.g.,][]{Fabbiano2}. Encouraged by this relation, we studied the colour properties of the optical counterparts of the point-like sources in our sample. In Table $\ref{t6}$, we classify our bright X-ray sources (log f$_{X}$ $\geq$ -13.5 erg cm$^{-2}$ s$^{-1}$) by their optical colour bi-modality by using the g-r/r or B-R/R methods (see section $\ref{s3-3}$). We used these parameters to get indications on the nature of X-ray emission in our galaxies. We found that the number of the red and blue galaxies is approximately equal (N$_{R}$ $\approx$ N$_{B}$) (see Fig. $\ref{1}$), and $\sim$ $\%55$ of the optical counterparts of the X-ray bright sources are identified as blue galaxies. On the other hand, a considerable number of our galaxies is found to be star-forming galaxies (see Fig. 3).  Because the most massive, short-lived, newly-formed stars can become high mass X-ray binaries (HMXB) that remain bright for $\sim$ 10$^{6-7}$ yr, the total X-ray emission closely tracks the star formation rate \citep[e.g.,][]{Helfand}. However, it is not possible to separate X-ray emission from HMXBs and LMXBs in distant galaxies. We note that low mass X-ray binary (LMXB) populations are quite low in late-type galaxies \citep[e.g.,][]{Grimm,Fabbiano}. In this case, the large number of the HMXBs might cause luminous X-ray emission (10$^{40}$ $<$ L$_{X}$ $<$ 10$^{42}$ erg s$^{-1}$) from these sources. However, we also note that high X-ray emission from late-type galaxies (L$_{X}$ $>$ $\times$ 10$^{41}$ erg s$^{-1}$) can also be produced by nuclear activity and a large population of XRBs at the same time. In some cases, supernova remnants (SNRs) can make small contributions to X-ray emission at lower luminosities. 
 
\subsection{Galaxy evolution within environment}
\begin{figure*}
\includegraphics*[width=8.3cm,angle=0]{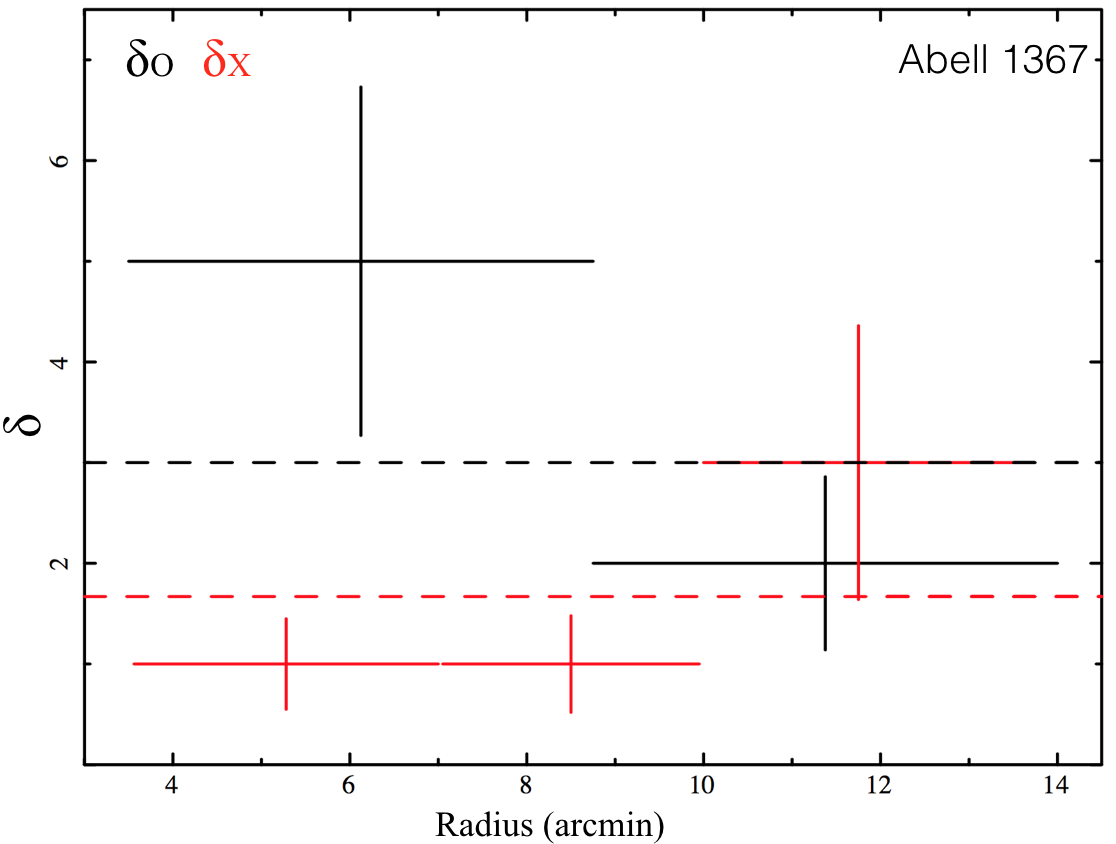}
\includegraphics*[width=8.35cm,angle=0]{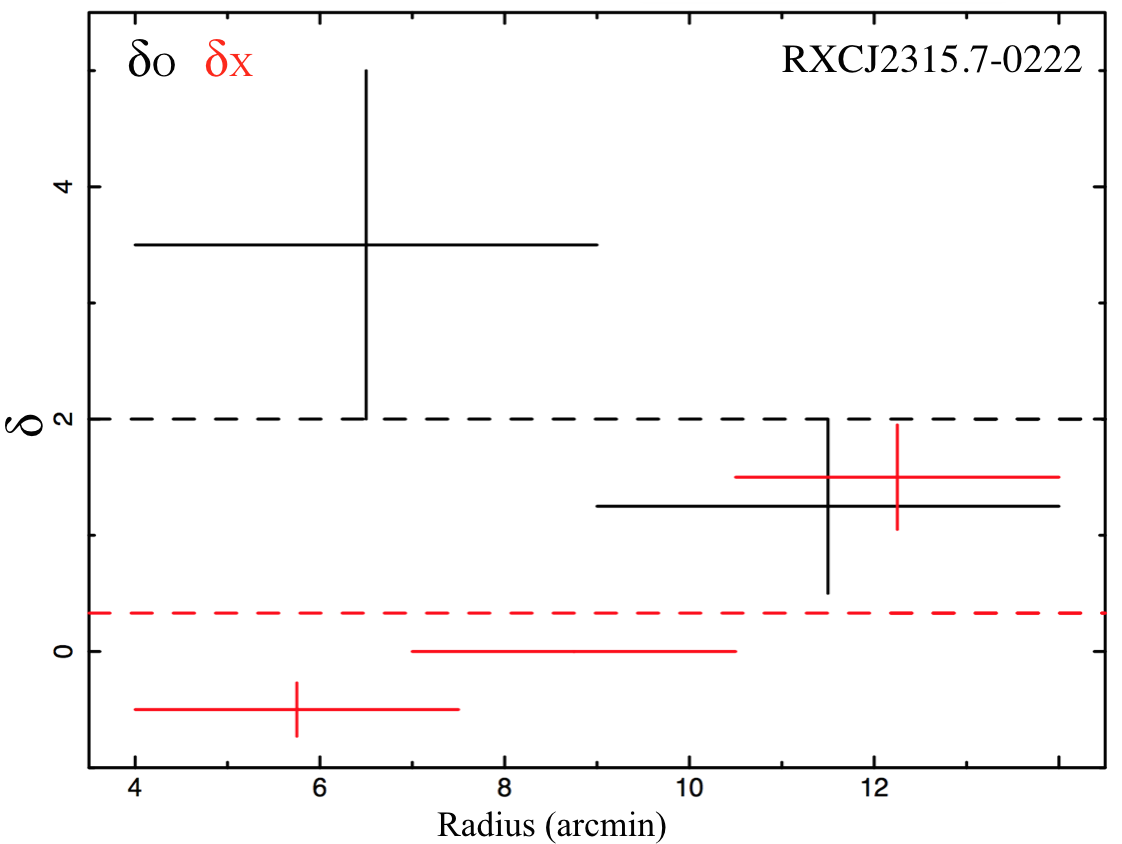}
\caption{X-ray versus optical overdensity as a function of the distance from the centre of the cluster. The red dashed line corresponds to the mean X-ray overdensity of relative galaxy cluster, and the black dashed line represents optical overdensity of relative galaxy cluster.}
\label{4}
\end{figure*}

\begin{figure}
\includegraphics*[width=7.45cm,angle=0]{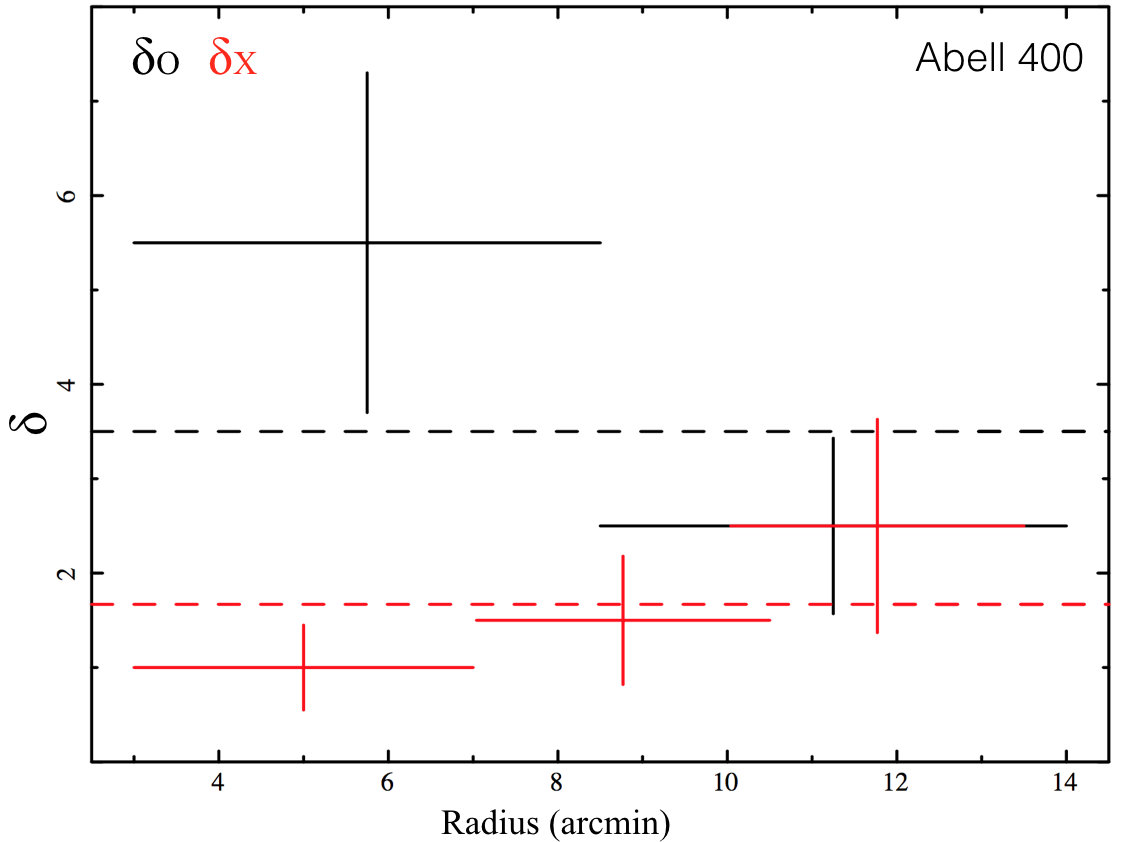}
\includegraphics*[width=7.45cm,angle=0]{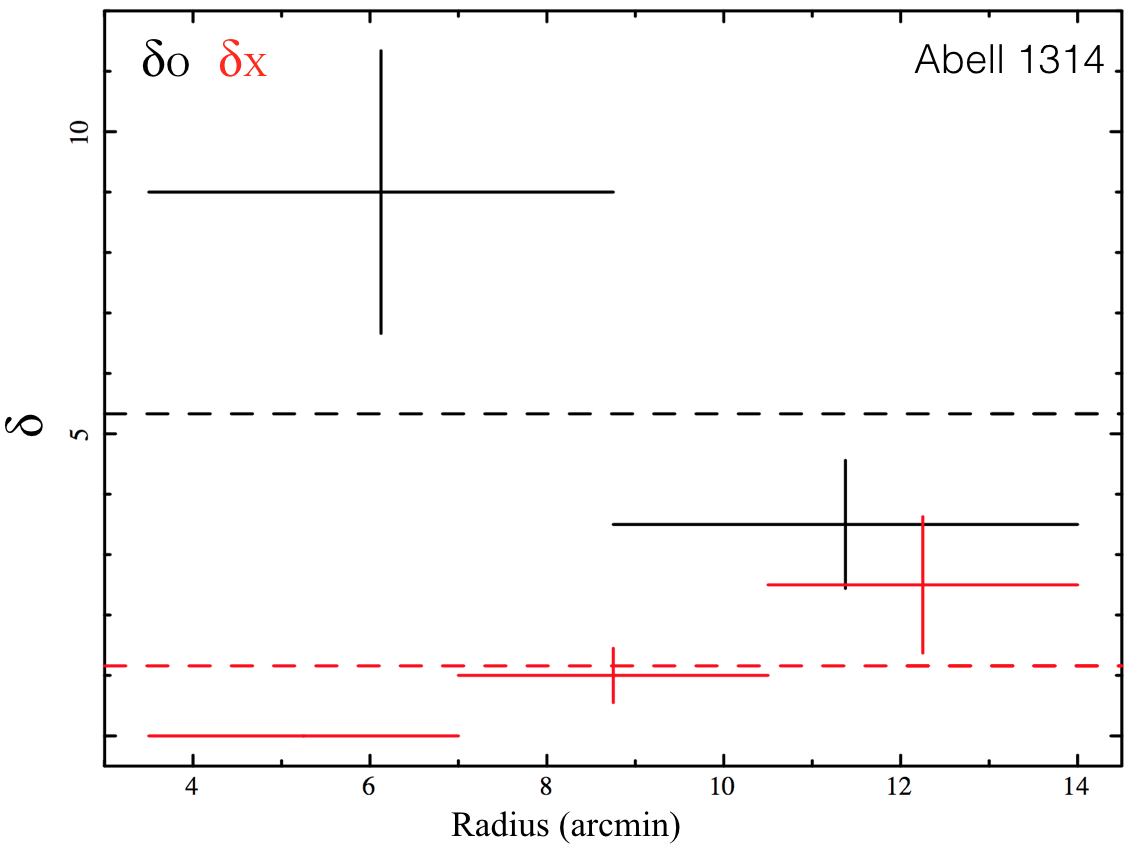}
\includegraphics*[width=7.45cm,angle=0]{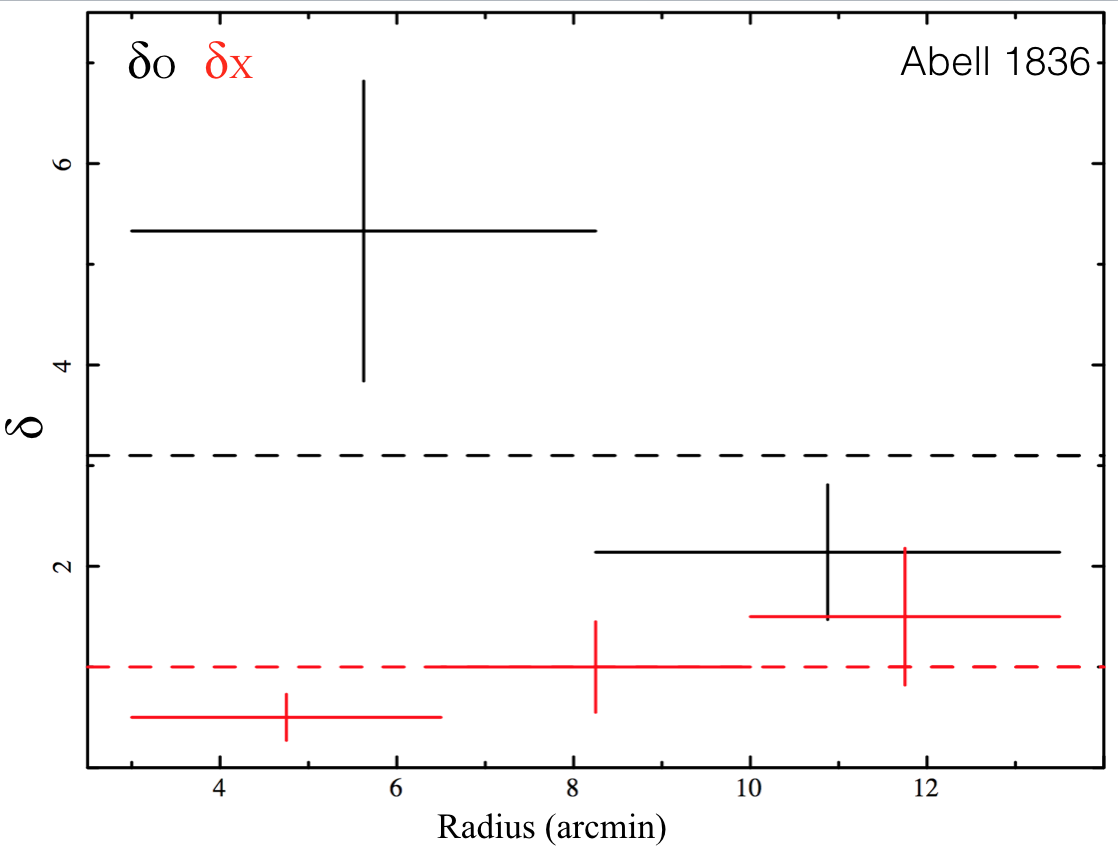}
\includegraphics*[width=7.45cm,angle=0]{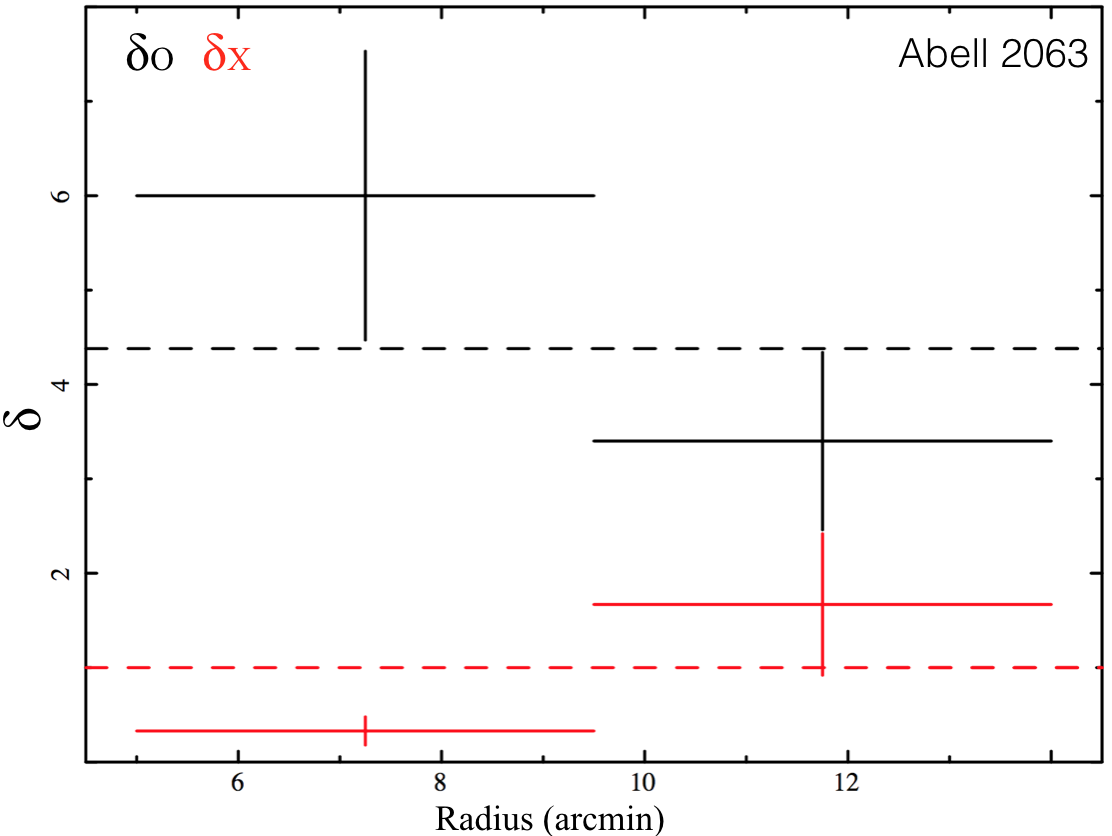}
\caption{X-ray versus optical overdensity as a function of the distance from the centre of the cluster. The red dashed line corresponds to the mean X-ray overdensity of relative galaxy cluster, and the black dashed line represents optical overdensity of relative galaxy cluster.}
\label{5}
\end{figure}

\begin{figure}
\includegraphics*[width=7.45cm,angle=0]{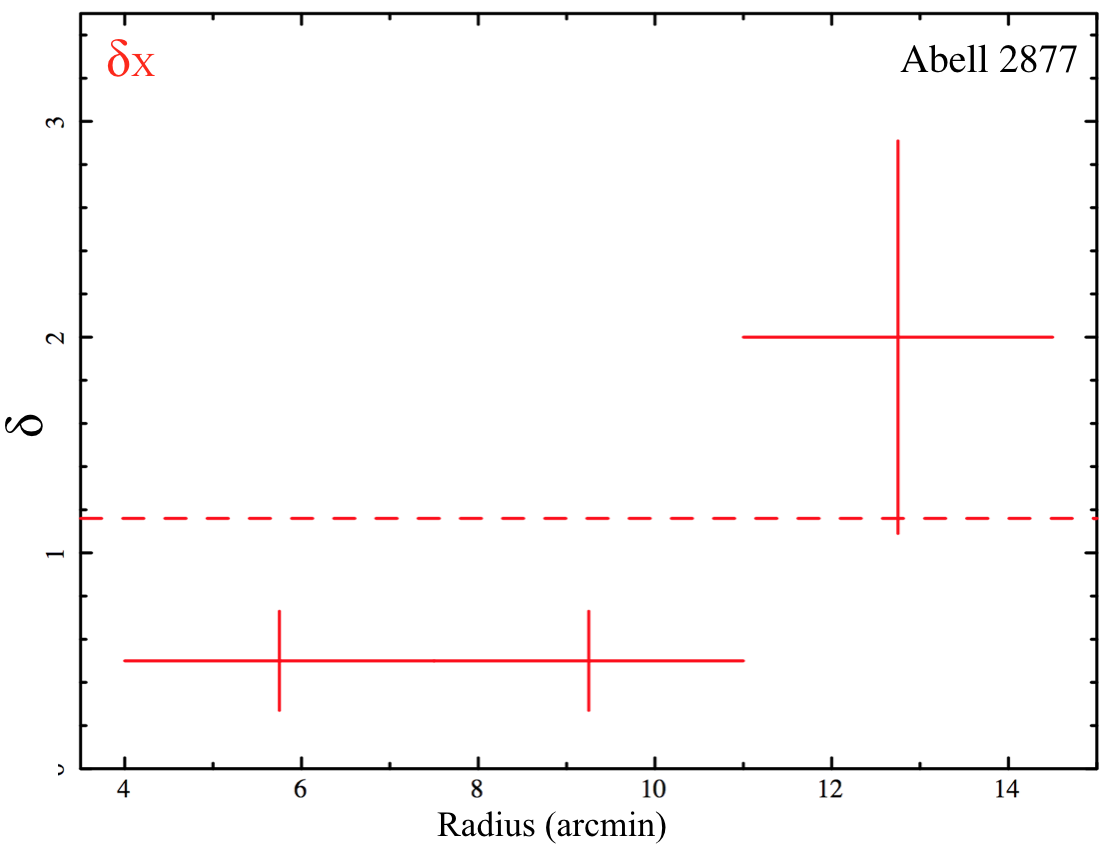}
\includegraphics*[width=7.45cm,angle=0]{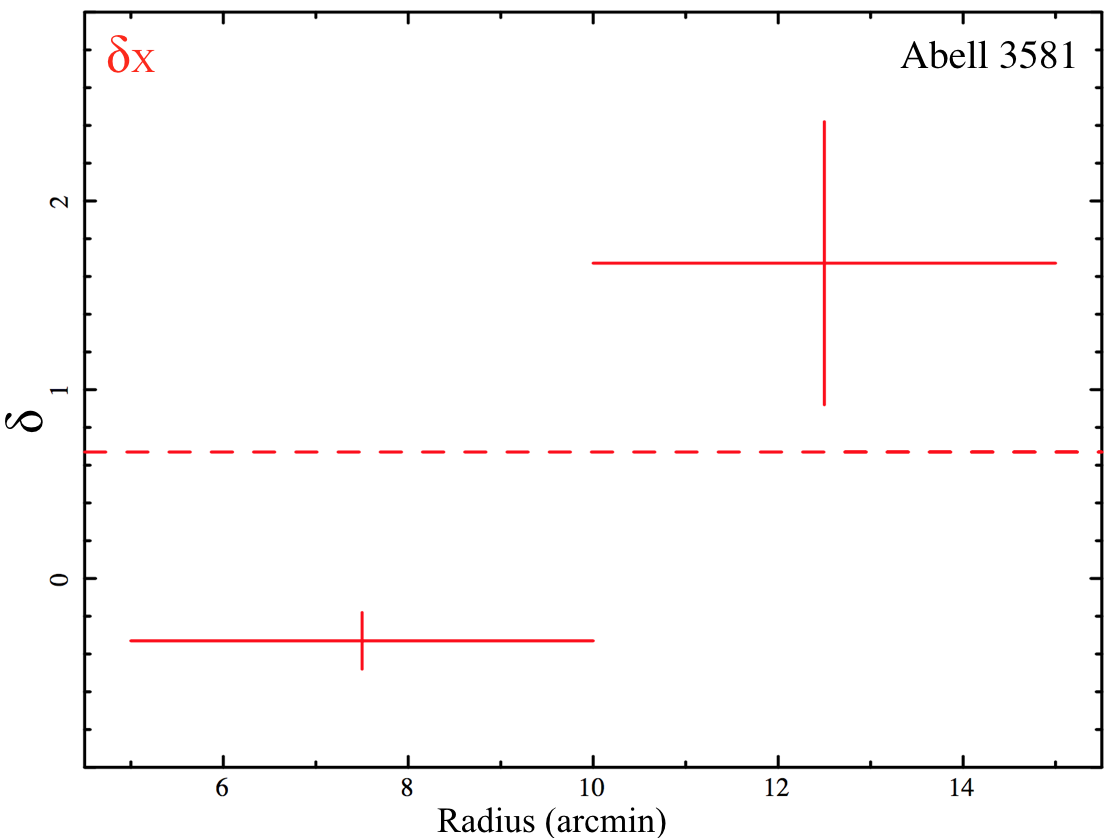}
\includegraphics*[width=7.45cm,angle=0]{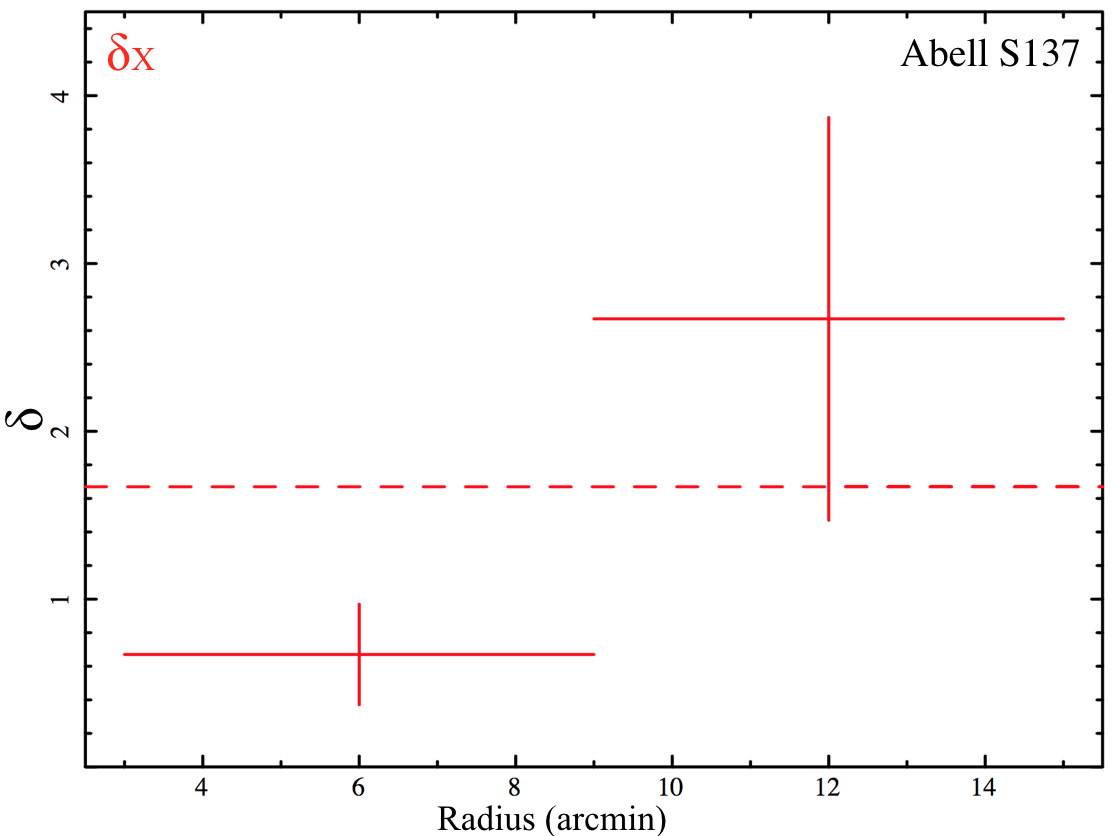}
\includegraphics*[width=7.45cm,angle=0]{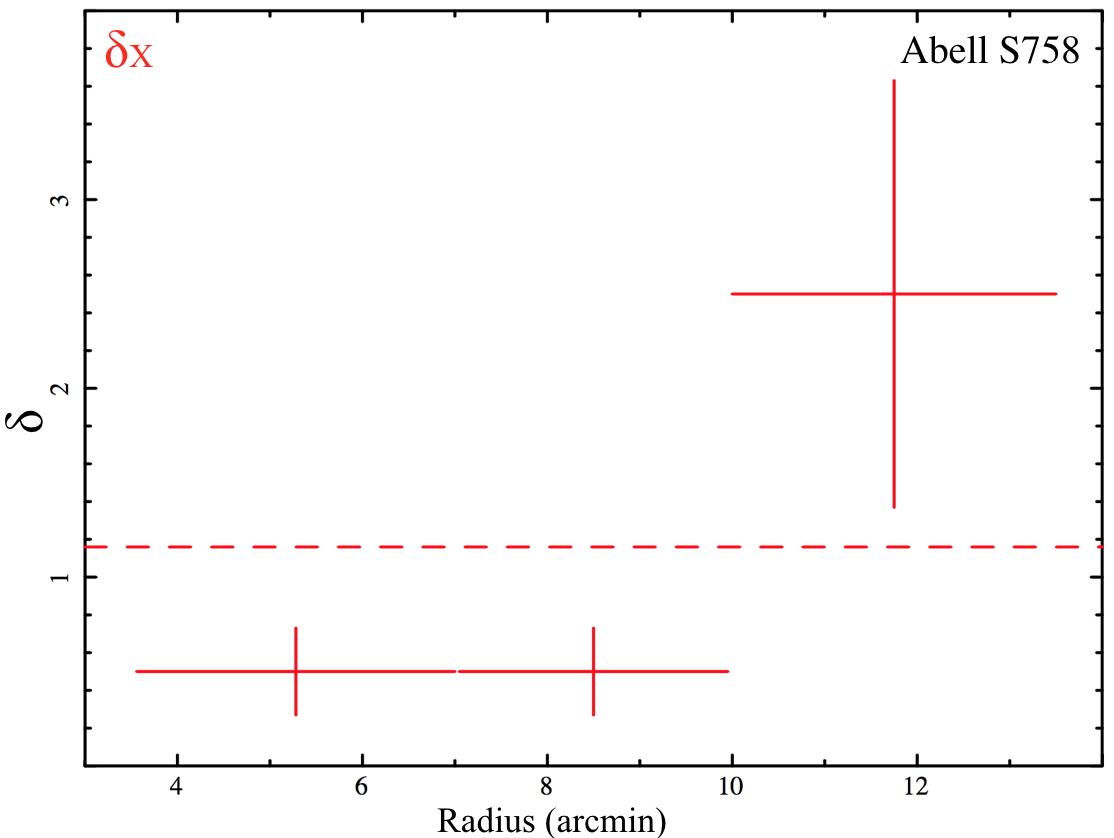}
\caption{X-ray versus optical overdensity as a function of the distance from the centre of the cluster. The red dashed line corresponds to the mean X-ray overdensity of relative galaxy cluster, and the black dashed line represents optical overdensity of relative galaxy cluster.}
\label{6}
\end{figure}

We studied X-ray overdensities from galaxy clusters relative to non-clustered fields. Expected X-ray source number densities were calculated in the Hubble Deep Field North, where number densities from fields are consistent with other field samples (see Fig. $\ref{2}$). We used SDSS archival data to obtain optical overdensities for 6 of the clusters in our sample, however, there are no SDSS observations for the remaining number of clusters. Optical galaxy overdensities were calculated in two divided areas by using characteristic magnitude method described in detail (see section $\ref{s4}$). X-ray and optical overdensities were compared with each other to address the nature of point-like X-ray emission. As a result, X-ray overdensities are found to be significantly lower than optical overdensities in our calculations (see Table $\ref{t3}$). However, we also point out that the X-ray overdensity of A1367 surprisingly reaches the mean optical overdensity at the outskirts of the cluster. This cluster shows an elongated shape through NW-SE direction, and two groups of star-forming galaxies are falling into the cluster's centre \citep{Cortese}. Recent studies imply that increased galaxy X-ray emission from cluster's field is probably caused by occurring merger events. \citet{Neal} reported triggered AGN activity from A2255 due to a cluster-cluster merger. Also, \citet{Hwang} studied two merging galaxy clusters and reported that cluster member galaxies show increased X-ray emission that can be related to both star formation and AGN activity. Therefore, member galaxies of A1367 are possibly triggered by ongoing merger events or in-falling of X-ray bright object that probably increased X-ray overdensity at the outskirts of A1367. In Figs $\ref{4}$, $\ref{5}$, and $\ref{6}$ we demonstrate the X-ray to optical overdensity comparisons as a function of radius. As can be seen from these figures, the optical galaxy densities decrease through outskirts of our clusters, whereas X-ray overdensities increase through outskirts. Our results reveal that X-ray sources are suppressed within the cluster environment, and suppression of X-ray AGNs increases through cluster's centre.  

\section{NOTES ON POINT-LIKE SOURCES}
In this section, we will present the results of an analysis meant to identify the main X-ray emission process (i.e., AGN or star formation) in a sub-sample of point-like sources. We present optically identified cluster members in Table $\ref{t5}$. In Table $\ref{t6}$, we present X-ray/optical properties of point sources by assuming them as likely cluster member. At low X-ray luminosities (10$^{41}$ $<$ L$_{X}$ $<$ 10$^{42}$ erg s$^{-1}$), it is not clear whether X-ray emission comes from SF or LLAGN. Therefore, measurement of SFR by using different methods can be very effective to resolve the nature of X-ray emission. We calculated the SFR of point sources by using  the equations defined by \citep{Condon1992} and \citep{Ranalli} respectively:

\begin{equation}
SFR [M\odot y^{-1}] = 2.5 \times 10^{-29} \times L_{1.4GHz} (erg  s^{-1} Hz^{-1}) 
\end{equation}

\begin{equation}
SFR [M\odot y^{-1}] = 2 \times 10^{-40} \times L_{2-10 keV} (erg  s^{-1})
\end{equation}

\subsection{The member galaxies}
\textbf{3C 75A} is identified as a pair member of NGC 1128 and the brightest galaxy of A400 \citep{LM}. This galaxy is elliptical and emits radio frequencies from relativistic jets \citep{BP}. Spectrum of this source contains absorption and thermal emission lines, and the best fit parameters are $\Gamma$=1.70$\pm{0.11}$, $norm_{pow}$= (8.28$\pm{2.52}$) $\times$ $10^{-6}$ $cm^{-5}$, nH=0.52$\pm{0.18}$ $cm^{-2}$, kT=0.8$\pm{0.05}$ keV, and $norm_{apec}$ = (6.76$\pm{2.18}$) $\times$ $10^{-5}$ $cm^{-5}$. Besides, X-ray hardness ratio is measured as -0.42$\pm{0.08}$. Furthermore, X-ray to optical comparison reveals a bright X-ray emission (log L$_{X}$ = 41.25) from this source. By considering all these facts, we predict that X-ray emission mostly comes from central AGN.
\textbf{NGC 3860} is identified as a strong AGN (possibly triggered by super-massive black hole) by \citet{Gavazzi}. We calculated \textit{X/O} = -1.64, log(L$_{X}$/L$_{B}$) = 30.68 and log(L$_{X}$/L$_{K}$)=30.23. Fitting the X-ray spectrum with a ($\Gamma$ = 1.30$\pm{0.14}$), fixed column density (nH = 1.82 $\times$ 10$^{20}$ cm$^{-2}$) and redshift (z=0.018663), we found a hardness ratio (HR) = -0.44$\pm{0.17}$ and log L$_{X}$ = 41.17 erg s$^{-1}$. According to these results, we claim that the X-ray emission process in NGC 3860 is due to nuclear activity, even though the source is an LLAGN.
\textbf{NGC 3862} is classified as brightest cluster galaxy (BCG) \citep{Sun} and AGN \citep{Veron,Gavazzi} due to its optical properties. Our results also reveal possible low luminous AGN activity (log L$_{X}$ = 41.76 erg s$^{-1}$) from this source (see Table $\ref{t5}$). 
\textbf{MCG+08-21-065} is a spiral galaxy (Sb)\citep{Nair}, and radio source (NVSS J113543+490214) is associated with this galaxy \citep{Condon}. Our analysis reveals very bright X-ray emission (log L$_{X} = $42.15 erg s$^{-1}$) from this source. Therefore, MCG+08-21-065 is an AGN. \citet{Shirazi} studied the optical spectrum of \textbf{2MASX J11340896+4915162} and classified this source as an AGN. On the basis of our results, we calculated log L$_{X}$ = 42.28 erg s$^{-1}$ and high X-ray to optical flux ratio (\textit{X/O} = -0.22) for this source. We confirm this source as AGN.  
\textbf{2MASX J15231224+0832590} is identified as a spiral galaxy (Sa) \citep{Leaman}. Spectral analysis of this source results in a $\Gamma$ = 1.63$\pm{0.12}$, logarithmic X-ray luminosity log L$_{X}$ = 42.40 erg s$^{-1}$, and hardness ratio (HR) = 0.35$\pm{0.04}$. In considering these results, we identify this source as an AGN. 
  \textbf{ESO 510- G 066} is identified as a lenticular galaxy (Sa0) \citep{Vaucouleurs}, and show shreds of evidence of radio jets \citep{Velzen}. Spectral analysis of this source demonstrated that this source is an unabsorbed X-ray source, where $\Gamma$ = 2.33$\pm{0.4}$. The investigation reveals enhanced X-ray emission with high X-ray to optical flux/luminosity ratio (\textit{X/O} = -1.88). Furthermore, log L$_{X}$ = 41.24 erg s$^{-1}$ and hardness ratio = -0.75$\pm{0.03}$ were calculated for this source. This galaxy is located in the outskirt of the A3581, and the X-ray centroid has a positional offset ($\sim$ 1.5 kpc) relative to the optical centroid. We calculated the star formation rate = 34.76 M$_{\odot}$/yr for a given X-ray luminosity of 41.24. This SFR measurement is in agreement with the one from the 1.4 GHz flux (SFR = 36.98 M$_{\odot}$/yr). By considering all these facts, we classify the source as a star-forming galaxy.
\textbf{NGC 3860B} is a spiral galaxy (S) and classified as HII region-like galaxy \citep{Gavazzi}. As expected from H II region-like galaxies, this source appears to emit UV emission \citep{Marcum}. In addition, \citet{Thomas} studied the H$_{\alpha}$ properties of this galaxy and reported SFR = 2.0 M$_{\odot}$/yr. The SFR values we computed using the radio and X-ray luminosities are in good agreement with the UV measurements, being 3.47 M$_{\odot}$/yr and 4.48 M$_{\odot}$/yr respectively. In conclusion, NGC 3860B appears to be X-ray normal galaxy.    
\subsection{New LLAGN candidates}
In this section, we concentrate on identifying new possible low luminous AGNs from our survey. To define LLAGNs, we studied X-ray properties, \textit{X/O}, galaxy colour and hardness ratio of point sources. Hardness ratio is defined as (H-S)/(H+S), where H is count rate in 2.0-10.0 keV band and S is count rate in 0.5-2.0 keV band. We present seven new LLAGN candidates in Table $\ref{t4}$. LLAGN selection is performed using following indicators:  \\
\\
- \textit{X/O} (> -1) \\
- Galaxy colour (Red) \\ 
- Hardness ratio ($>$ -0.55) \\
- Total X-ray counts ($>$ 100 cts) \\
- X-ray luminosity ($>$ 10$^{41}$ erg s$^{-1}$) \\

These indicators are very efficient to identify X-ray AGNs, and similar methods were applied to other AGN candidates on different surveys \citep[e.g.,][]{Xue, Ranalli2012,Vattakunnel,Marchesi}. 

\begin{table}
\begin{center}
  \resizebox{0.53\textwidth}{!}{\begin{minipage}{0.62\textwidth}
        \caption{X-ray to optical properties of new LLAGN candidates.}
  \begin{tabular}{cccccc}
\hline
\hline
Source Name  & HR & log(L$_{X}$)  & \textit{X/O} & Cluster \\
& & erg s$^{-1}$ & & \\
\hline
\vspace{0,5mm}
XMMU J140721.6-264716 & 0.62$\pm{0.11}$  & 41.30 & -0.68 & A3581 \\
\vspace{0,5mm}
XMMU J113408.4+490318 & -0.54$\pm{0.07}$ & 41.24 & -0.02 & A1314 \\
\vspace{0,5mm}
XMMU J140215.6-113748 & -0.07$\pm{0.05}$  & 41.84 & 0.21 & A1836 \\
\vspace{0,5mm}
XMMU J011105.5-612548 & 0.29$\pm{0.11}$  & 41.16 & -0.61 & AS137 \\
\vspace{0,5mm}
XMMU J010949.4-613153 & -0.33$\pm{0.09}$  & 41.13 & 0.02 & AS137 \\ 
\vspace{0,5mm}
XMMU J141216.6-342422 & -0.45$\pm{0.04}$ & 41.46 & -0.10 & AS758 \\
\vspace{0,5mm}
XMMU J141308.6-342105 & 0.30$\pm$0.22 & 41.49 & -0.96 & AS758 \\
\hline
\label{t4}
\end{tabular}
\end{minipage}}
\end{center}
\end{table}
\section{CONCLUSIONS}
In this work, we studied ten nearby ($\leq$ 171 Mpc) galaxy clusters. Within these clusters, we detected 874 point-like sources; a fraction of them (483) is expected to be a false detection related to the diffuse ICM emission. We removed those sources located in the central regions of galaxy clusters (95 $\leq$ r$_{c}$ $\leq$145 kpc) from our final sample unless they are bright enough to be detected within ICM. All the point-like sources within 0.3-10 keV spectra were fitted with an absorbed power-law; a minority of spectra showed evidence of thermal emission lines, which we fitted adding a thermal component (APEC).  We calculated the log \textit{N} - log \textit{S} for our samples and we compared cluster results with those obtained in the Lockman Hole and in the Hubble Deep Field North. The number counts are a factor $\sim$ 2 higher in the clusters than they are in the fields, at any flux level. In the luminosity range (40.08 $\leq$ log(L$_{X}$) $\leq$ 42.39 erg s$^{-1}$) of the point-like sources in our sample, X-ray emission is mostly produced from LLAGNs, XRBs and star formation. Although starburst and normal galaxies dominate large fraction of X-ray sources of our survey, the fraction of LLAGNs is nonetheless significant. Using proxies such as \textit{X/O}, \textit{L$_{X}$/L$_{B}$} and \textit{L$_{X}$/L$_{K}$}, we found significant X-ray excess in several galaxies. By considering X-ray excess of member galaxies, we linked the nature of X-ray emission to two different processes: AGN triggering and star formation. We used efficient indicators to separate LLAGNs and star-forming galaxies. In the majority of the red galaxies, the enhanced X-ray emission can be explained by AGN activity; nevertheless X-ray emission can be produced by unresolved XRBs in some cases. For the blue galaxies, we explained X-ray excess with star formation, which can be related to an extreme number of HMXBs and/or contributions from SNRs. Due to the absence of redshift information of X-ray sources, we assumed all X-ray sources in our survey as cluster members, and we compared X-ray and optical overdensities of our sample of clusters. We found that X-ray overdensities are significantly lower than optical overdensities in our survey, which can be explained by the fact that X-ray sources are suppressed within cluster environments. We also note that some non-redshift X-ray sources may not be cluster members. In that case, calculated X-ray overdensities may decrease, and suppression of X-ray sources in cluster environments even becomes clearer. The absence of very bright X-ray sources (L$_{X}$ $>$ 10$^{42}$ erg s$^{-1}$) in nearby galaxy clusters indicates that X-ray AGNs are the highly suppressed within the central regions of clusters due to highly pressurised environment. We still note that although dense and hot ICM suppress X-ray AGNs, AGN fuelling can still be effective in the sparse parts of ICM. As possible as this scenario is, we conclude that the large majority X-ray bright galaxies at the outskirts of clusters are dominated by star formation activities. Furthermore, we explain the existence of LLAGNs within clusters with close encounters of galaxies rather than AGN fuelling. Consequently, we contributed the suppression/triggering conflict in favour of the suppression by studying ten nearby galaxy clusters. However, the number of clusters in our sample is quite low, and more SDSS and XMM-Newton observations of nearby galaxy clusters are required to solve the conflict.

\section*{Acknowledgements}

We are grateful to the anonymous referee for comments that significantly improved this article. We would like to thank Guenther Hasinger, Marat Gilfanov, Ho Seong Hwang, Elias Koulouridis, Piero Ranalli and Stefano Marchesi for their valuable comments and suggestions. We acknowledge the financial support provided by The Scientific and Technological Research Council of Turkey through grant no: 113F117. The authors also would like to thank YTU Scientific Research \& Project Office (BAP) funding with contact number 2013-01-01-KAP04. 








\clearpage 

\appendix

\section{X-ray to optical properties of galaxies}

X-ray and optical properties of identified member galaxies I) Galaxy names II) Redshift values from Ned Astronomical Database III) R-band magnitude values from Vizier database IV) X-ray to optical flux ratio V) Apparent blue magnitude values from Vizier database VI) Apparent k magnitude values are taken from \citet[][]{Tully,Dalya} VII) Hard band logarithmic X-ray luminosity values from spectral analysis VIII) K-band luminosities calculated from extinction corrected k-band magnitude. IX) Logarithmic blue optical luminosity values calculated from extinction corrected b-band magnitudes X) Morphological type of galaxies XI) Name of cluster hosts identified galaxies.

\begin{table}
\caption{X-ray to optical properties of individual galaxies.}
\vspace{2.5mm}
\centering
\scalebox{0.83}{
\begin{tabular}{ccccccccccc}
\hline
\hline
 Object Name & Redshift & m$_{r}$ & \textit{X/O} & m$_{B}$ & m$_{K}$ & log L$_{X}$ & log(L$_{K}$/L$_{\odot}$) & log (L$_{B}$/L$_{\odot}$) & Type & Cluster  \\

 (I) & (II) & (III) & (IV) & (V) & (VI) & (VII) & (VIII) & (IX) & (X) & (XI) \\
 \hline
 \hline

CGCG 415-040 & 0.022980 & 14.35 & -2.19 & 14.53 & 10.34 & 40.70 & 11.15 & 10.33 & S0 & A400 \\

CGCG 415-046 & 0.022820 & 14.50 & -2.28 & 14.48 & 10.79 & 40.55 & 10.96 & 10.34 & E & A400 \\

3C 75A & 0.022580 & 14.99 & -1.48 & 13.86 & 9.32 & 41.25 & 11.53 & 10.56 & E & A400 \\

3C 75B & 0.024113 & 13.10 & -2.66 & 15.00 & 12.15 & 40.71 & 10.46 & 10.17 & S0 & A400 \\

2MASX J02574741+0601395 & 0.024811 & 15.09 & -1.93 & 15.72 & 11.29 & 40.67 & 10.83 & 9.91 & E & A400 \\

2MASX J01100662-4555544 & 0.024360 & 15.46 & -1.38 & 16.14 & 12.42 & 41.05 & 10.38 & 9.74 & S0 & A2877 \\

2MASX J01101993-4551184 & 0.023243 & none & none & 15.31 & 10.61 & 40.84 & 11.06 & 10.03 & S0 & A2877 \\

IC1633 & 0.024240 & 12.95 & -2.14 & 12.40 & 8.39 & 41.85 &11.98 & 11.23 & E1 & A2877 \\

ESO 243- G 049 & 0.022395 & 14.29 & -1.77& 14.77 & 10.70 & 41.08 & 10.99 & 10.21 & Sa0 & A2877 \\

ESO 243- G 051 & 0.021855 & 14.00 & -2.68 & 13.62 & 10.01 & 40.36 & 11.23 & 10.64 & Sb & A2877 \\

ESO 243- G 045 & 0.025881 & 13.81 & -2.25 & 13.32 & 9.71 & 40.87 & 11.51 & 10.92 & S0 & A2877 \\

NGC 3851 & 0.021130 & 14.41 & -2.29 & 15.12 & 11.01 & 40.48 & 10.89 & 10.10 & E & A1367 \\

NGC 3860 & 0.018663 & 14.32 & -1.64 & 13.77 & 10.39 & 41.17 & 10.99 & 10.49 & Sa & A1367 \\

NGC 3860B & 0.028250 & 15.69 & -1.91 & 14.99 & 13.35 & 40.35 &10.17 & 10.37 & S & A1367 \\

NGC 3861 & 0.016900 & 13.88 & -2.42 & 12.93 & 9.95 & 40.56 &11.12 & 10.78 & S & A1367 \\

NGC 3862 & 0.021718 & 13.64 & -1.37& 13.51 & 9.48 & 41.76 & 11.52 & 10.76 & E & A1367 \\

CGCG 097-125 & 0.027436 & 15.06 & -1.77 & 15.23 & 11.46 & 40.74 & 10.92 & 10.27 & E & A1367 \\

NGC 3842 & 0.021068 & 12.18 & -3.12 & 12.62 & 9.07 & 40.55	 & 11.65 & 11.09 & E & A1367 \\

GALEXASC J114359.29+195633.6 & 0.023323 & 19.74 & -0.08 & 21.38 & none  & 40.56	& none & 7.67 & & A1367  \\

2MASX J15225650+0839004 & 0.03361 & 15.00 & -2.09 & 15.84 &11.91  & 40.86 & 10.90 & 10.18 & S0 & A2063 \\

CGCG 077-097 & 0.034174 & 13.14 & -1.96 & 14.36	 & 10.07  & 41.72 & 11.65 & 10.79 & S? & A2063 \\

2MASX J15231224+0832590 & 0.036619 & 15.20 & -0.49 & 15.20 & 12.41  & 42.39 & 10.78 & 10.51 & Sa & A2063 \\

MCG -02-36-002 & 0.037776 & 12.74 & -2.61 & 14.10 & 9.97  & 41.31 & 11.79 & 10.99 & Sa0 & A1836 \\

2MASX J14015570-1138043  & 0.036979 & 16.57 & -1.79 & 18.15 & 14.09  & 40.59 & 10.12 & 9.35 & & A1836 \\

2MASX J14013206-1139261  & 0.041662 & 15.55 & -1.74 & 15.60 & 11.86 & 41.12 & 11.12 & 10.47 & & A1836 \\

IC 708 & 0.031679 & 13.07 & -2.54 & 13.85 & 10.09  & 41.09 & 11.58 & 10.93 & E & A1314 \\

IC 711  & 0.032436 & 13.88 & -2.59 & 14.88 & 11.08  &40.72 & 11.21 & 10.54 & E? & A1314 \\

IC 712  & 0.033553 & 13.13 & -2.98 & 14.05 & 9.89  & 40.68 & 11.71 & 10.90 & S? & A1314 \\

2MASX J11340896+4915162 & 0.037230 & 16.25 & -0.22 & 16.99 & 13.47  & 42.28 & 10.33 & 9.78 & & A1314 \\

MCG+08-21-065 & 0.029670 & 15.14 & -0.58 & 15.21 & 11.88 & 42.15 & 10.77 & 10.29 & Sb & A1314 \\

LEDA 97398 & 0.031600 & 19.70 & -0.38 & 20.80 & none & 40.65 & none & 8.11 & & A1314 \\  

IC 4374 & 0.021798 & 13.79 & -1.44 & 15.24 & 9.54 & 41.58 & 11.49 & 10.06 & Sa0 & A3581 \\

ESO 510- G 065 & 0.025671 & 14.21 & -2.76 &  16.11 & 11.90  & 40.24 & 10.68 & 9.85 & Sb & A3581 \\
 
ESO 510- G 066 & 0.024333 & 13.78 & -1.88 & 15.13 & 9.92 & 41.24 & 11.43 & 10.19 & Sa0 & A3581 \\

MCG-06-31-029 & 0.038500 & 11.10 & -3.89 & 11.20 & 11.15  & 40.72 & 11.34 &12.17 & E+ & AS758 \\

2MASX J14122917-3417417 & 0.043003 & 12.41 & -3.49 &  15.36 & 11.24 & 40.67 & 11.39 & 10.60 & & AS758 \\

NGC 0432 & 0.026929 & 13.82 & -2.89 & 13.92 & 9.93  & 40.29 & 11.47 & 10.72 & S0 & AS137 \\

2MASX J01125179-6139513 & 0.026442 & 14.05 & -2.65 & 14.75 & 10.70 & 40.43 & 11.15 & 10.38 & E & AS137 \\

NGC 7556 & 0.025041 & 12.21 & -3.23 & 15.39 & 9.24 & 40.56 &  11.65 & 10.04 & S0 & RXCJ2315.7-0222 \\

NGC 7566 & 0.026548 & 13.03 & -3.12 & 13.66 & 10.19  & 40.32 & 11.33 & 10.79 & Sb? & RXCJ2315.7-0222 \\

\vspace{0.5mm}
\label{t5}
\end{tabular}
}
\end{table}

\clearpage

\section{Galaxy Colour - X-ray Point-like Emission Relation}
I) Source name II) Hardness Ratio [defined as (H-S)/(H+S), where H is count rate in 2.0-10.0 keV and S is count rate in 0.5-2.0 keV] III) Logarithmic X-ray flux IV) Logarithmic X-ray luminosity V) R-band magnitude VI) G-R/B-R Values VII) Projected distance from centre of related galaxy cluster VIII) Likelihood of optical association IX) Galaxy type in colour X) Cluster name.

\begin{table}
\caption{Galaxy Colour Survey}
\vspace{2.5mm}
\centering
\scalebox{0.92}{
\begin{tabular}{cccccccccc}
\hline
Object Name & HR & Log f$_{X}$ & Log L$_{X}$ & r & g-r & d & P & Type & Cluster \\
 &  & erg cm$^{-2}$ s$^{-1}$ & erg s$^{-1}$ & mag & mag & Mpc & $\%$ & & \\
 I & II & III & IV & V & VI & VII & VIII & IX & X \\
\hline
XMMU J025824.6+060248 & -0.61$\pm$0.05 & -12.72 & 41.40 & 18.58 & 0.33 & 0.31 & 84 & Blue & A400 \\
XMMU J025749.2+055136 & -0.49$\pm$0.09 &-12.95 & 41.18 & 21.09 & 0.02 & 0.29 & 98 & Blue & A400 \\
XMMU J025747.3+060942 & -0.54$\pm$0.13 & -13.43 & 40.70 & 20.67 & -0.04 & 0.24 & 91 & Blue & A400 \\
XMMU J025718.5+060022 & -0.57$\pm$0.15 & -13.34 & 40.76 & 20.73 & 0.59 & 0.17 & 92 & Blue & A400 \\
XMMU J025730.8+060545 & -0.34$\pm$0.20 & -13.44 & 40.69 & 18.53 & 0.64 & 0.15 & 97 & Blue & A400 \\
XMMU J025718.8+060813 & -0.45$\pm$0.21 & -13.48 & 40.65 & 17.60 & 0.84 & 0.25 & 96 & Red & A400 \\
XMMU J025806.2+055327 & 0.09$\pm$0.32 & -13.31 & 40.82 & 20.91 & 0.02 & 0.29 & 95 & Red & A400 \\
XMMU J025712.6+055960 & 0.88$\pm$0.17 & -13.42 & 40.71 & 21.07 & 0.66 & 0.21 & 96 & Red & A400 \\
XMMU J113421.6+490050 & -0.60$\pm$0.08 & -13.47 & 40.95 & 19.47 & 0.66 & 0.24 & 98 & Blue & A1314 \\ 
XMMU J113408.4+490318 & -0.54$\pm$0.11 &-13.16 & 41.24 & 19.09 & 1.01 & 0.28 & 83 & Red & A1314 \\
XMMU J113548.5+491150 & -0.44$\pm$0.24 &-13.48 & 40.93 & 20.95 &  0.48 & 0.49 & 95 & Blue & A1314 \\
XMMU J114435.0+195131 & -0.58$\pm$0.08 & -13.07 & 40.97 & 20.01 & 0.65 & 0.20 & 94 & Blue & A1367 \\
XMMU J114427.1+194338 & -0.62$\pm$0.18 & -13.32 & 40.72 & 21.09 & 0.45 & 0.09 & 89 & Blue & A1367 \\
XMMU J114452.5+195133 & -0.55$\pm$0.10 & -13.02 & 41.02 & 21.89 & 0.32 & 0.21 & 98 & Blue & A1367 \\
XMMU J114515.8+194951 & -0.61$\pm$0.28 & -13.38 & 40.66 & 19.84 & 0.27 & 0.28 & 97 & Blue & A1367 \\
XMMU J114436.5+195336 & -0.29$\pm$0.30 & -13.48 & 40.56 & 20.89 & 0.19 & 0.25 & 99 & Blue & A1367 \\
XMMU J114359.3+195632 & -0.73$\pm$0.21 & -13.48 & 40.56 & 21.75 & 0.70 & 0.30 & 73 & Red & A1367 \\
XMMU J114507.7+193552 & -0.82$\pm$0.37 & -13.49 & 40.55 & 21.27 & 1.09 & 0.31 & 92 & Red & A1367 \\
XMMU J140215.6-113748 & -0.07$\pm$0.10 & -12.63 & 41.84 & 18.36 & 1.05 & 0.37 & 96 & Red & A1836 \\
XMMU J140207.9-113553 & -0.63$\pm$0.08 & -13.42 & 41.06 & 19.53 & 0.11 & 0.28 & 94 & Blue & A1836 \\
XMMU J140135.5-112708 & -0.15$\pm$0.34 & -13.39 & 41.05 & 19.28 & 1.51 & 0.42 & 93 & Red & A1836 \\
XMMU J140043.8-113731 & -0.29$\pm$0.40 & -13.42 & 41.18 & 17.92 & 1.30 & 0.63 & 96 & Red & A1836 \\ 
XMMU J152342.5+084535 & -0.82$\pm$0.32 & -13.22 & 41.23 & 22.00 & 0.96 & 0.55 & 74 & Red & A2063 \\
XMMU J152250.9+084447 & 0.41$\pm$0.76 & -13.21 & 41.21 & 21.34 & 1.15 & 0.38 & 92 & Red & A2063 \\
XMMU J152322.3+082159 & -0.20$\pm$0.48 & -13.22 & 41.23 & 19.99 & 0.23 & 0.64 & 97 & Blue & A2063 \\
XMMU J231638.6-022527 & -0.56$\pm$0.07 & -12.54 & 41.67 & 19.05 & 0.20 & 0.42 & 99 & Blue & RXCJ2315.7-0222 \\
XMMU J231624.5-021457 & -0.77$\pm$0.07 & -13.35 & 40.80 & 20.63 & 1.57 & 0.40 & 81 & Red & RXCJ2315.7-0222 \\
XMMU J231555.9-021644 & 0.33$\pm$0.29 & -13.33 & 40.81 & 23.84 & 0.44 & 0.21 & 93 & Blue & RXCJ2315.7-0222 \\
XMMU J231610.3-021502 & -0.64$\pm$0.31 & -13.47 & 40.67 & 20.19 & 1.43 & 0.31 & 91 & Red & RXCJ2315.7-0222 \\
XMMU J231500.7-022515 & -0.25$\pm$0.27 &-13.36 & 40.85 & 18.72 & 0.72 & 0.34 & 97 & Blue &  RXCJ2315.7-0222 \\
\hline
Object Name & HR & Log f$_{X}$ & Log L$_{X}$ & R & B-R  & d & P & Type & Cluster \\
 & & erg cm$^{-2}$ s$^{-1}$ & erg s$^{-1}$ & mag & mag  & Mpc & $\%$ & & \\
  I & II & III & IV & V & VI & VII & VIII & IX & X \\
\hline
XMMU J010914.4-455914 & -0.61$\pm$0.09 & -13.02 & 41.13 & 18.98 & 0.74 & 0.23 & 93 & Blue & A2877 \\
XMMU J010952.4-460536 & -0.63$\pm$0.26 & -13.47 & 40.67 & 19.37  & 0.59 & 0.29 & 96 & Blue & A2877 \\
XMMU J011024.5-454426 & -0.82$\pm$0.27 &-13.44 & 40.70 & 18.97 & 0.48 & 0.36 & 98 & Blue & A2877 \\
XMMU J010858.9-455136 & -0.01$\pm$0.49 & -13.20 & 40.93 & 20.18 & 1.09 & 0.31 & 99 & Red & A2877 \\
XMMU J140721.6-264716 & 0.62$\pm$0.16 & -12.77 & 41.30 & 16.47 & 1.65 & 0.40 & 92 & Red & A3581 \\ 
XMMU J140803.6-270841 & -0.21$\pm$0.60 & -13.49 & 40.58 & 20.17 & 1.31 & 0.31 & 85 & Red & A3581 \\
XMMU J140825.4-270849 & -0.39$\pm$0.35 & -13.25 & 40.82 & 20.17 & 0.71 & 0.42 & 85 & Blue & A3581 \\
XMMU J011205.4-613255 & -0.56$\pm$0.09 & -13.47 & 40.73 & 20.16 & 0.77 & 0.08 & 99 & Blue & AS137 \\
XMMU J011127.4-612612 & -0.39$\pm$0.24 & -13.49 & 40.70 & 18.35 & 0.87 & 0.18 & 99 & Blue & AS137 \\
XMMU J011105.5-612548 & 0.29$\pm$0.20 & -13.02 & 41.16 & 17.28 & 1.63 & 0.24 & 88 & Red & AS137 \\
XMMU J010949.4-613153 & -0.33$\pm$0.16 & -13.06 & 41.13 & 18.95 & 1.57 & 0.44 & 62 & Red & AS137 \\
XMMU J011213.1-612015 & -0.43$\pm$0.20 & -13.36 & 40.84 & 19.34 & 0.49 & 0.37 & 99 & Blue & AS137 \\
XMMU J141117.5-341116 & -0.58$\pm$0.16 & -13.09 & 41.42 & 18.65 & 0.70 & 0.74 & 87 & Blue & AS758 \\
XMMU J141216.6-342422 & -0.45$\pm$0.06 & -13.06 & 41.46 & 18.65 & 1.70 & 0.19 & 93 & Red & AS758 \\
XMMU J141308.6-342105 & 0.30$\pm$0.22 & -13.01 & 41.49 & 16.37 & 1.58 & 0.46 & 93 & Red & AS758 \\
XMMU J141245.4-342343 & -0.46$\pm$0.10 & -13.35 & 41.18 & 18.45 & 0.82 & 0.28 & 87 & Blue & AS758 \\
XMMU J141223.0-341330 & -0.51$\pm$0.13 & -13.47 & 41.06 & 18.65 & 0.50 & 0.32 & 88 & Blue & AS758 \\
\hline
\label{t6}
\end{tabular}
}
\end{table}

\clearpage

\begin{table}
\caption{Properties of point sources studied in this work.}
\vspace{3mm}
\centering
\scalebox{0.92}{
\begin{tabular}{ccccccc}
\hline
\hline
Object Name & Net Counts & Log Flux & Log Luminosity & Cluster & Optical Counterpart & Redshift  \\
 & & erg cm$^{-2}$ s$^{-1}$ & erg s$^{-1}$ & & & \\
 \hline
 \hline
XMMU J025824.6+060248 & 750 & -12.72 & 41.40 & A400 & 2XMM J025824.6+060248 & \\
XMMU J025741.5+060136 & 570 & -12.98 & 41.25 & A400 & 3C 75A & 0.02258 \\
XMMU J025741.8+060120 & 343 & -13.40 & 40.71 & A400 & 3C 75B & 0.024113 \\ 
XMMU J025749.2+055136 & 431 & -12.95 & 41.18 & A400 & 2XMM J025749.2+055136 & \\
XMMU J025724.7+060156 & 481 & -13.21 & 40.88 & A400 & 2XMM J025724.7+060156 & \\
XMMU J025808.1+055808 & 233 & -13.34 & 40.76 & A400 &  & \\
XMMU J025747.3+060942 & 245 & -13.43 & 40.70 & A400 &  & \\
XMMU J025802.5+055448 & 221 & -13.47 & 40.67 & A400 &  & \\
XMMU J025802.3+055213 & 168 & -13.67 & 40.47 & A400 &  & \\
XMMU J025718.5+060022 & 258 & -13.34 & 40.76 & A400 &  & \\
XMMU J025736.7+060822 & 190 & -13.58 & 40.55 & A400 &  & \\
XMMU J025730.8+060545 & 197 & -13.44 & 40.69 & A400 &  & \\
XMMU J025718.8+060813 & 131 & -13.48 & 40.65 & A400 &  & \\
XMMU J025803.6+061107 & 102 & -13.49 & 40.64 & A400 &  & \\
XMMU J025806.2+055327 & 93 & -13.31 & 40.82 & A400 &  & \\
XMMU J025820.9+060008 & 107 & -13.65 & 40.48 & A400 &  & \\
XMMU J025747.1+060136 & 41 & -13.47 & 40.67 & A400 & 2MASX J02574741+0601395  & 0.024811 \\ 
XMMU J025712.6+055960 & 132 & -13.42 & 40.71 & A400 &  & \\
XMMU J025733.7+055835 & 244 & -13.43 & 40.70 & A400 &CGCG 415-040 & 0.02298 \\
XMMU J025716.6+055736 & 138 & -13.66 & 40.47 & A400 &  & \\ 
XMMU J025821.0+060537 & 120 & -13.58 & 40.55 & A400 & CGCG 415-046 & 0.02282 \\
XMMU J025752.1+060631 & 97 & -13.52 & 40.61 & A400 &  & \\
XMMU J025810.2+055948 & 148 & -13.51 & 40.62 & A400 &  & \\
XMMU J025812.7+055828 & 83 & -13.92 & 40.21 & A400 &  & \\
XMMU J025811.2+055227 & 78 & -13.39 & 40.74 & A400 &  & \\
XMMU J025824.2+055810 & 58 & -13.84 & 40.29 & A400 &  & \\
XMMU J025702.5+060945 & 63 & -13.85 & 40.28 & A400 &  & \\
XMMU J025709.9+060320 & 69 & -13.70 & 40.42 & A400 &  & \\
XMMU J025711.7+060160 & 92 & -13.77 & 40.36 & A400 &  & \\
XMMU J025751.7+054843 & 49 & -13.35 & 40.76 & A400 &  & \\
XMMU J025712.7+061114 & 60 & -13.81 & 40.32 & A400 &  & \\
XMMU J025807.0+060155 & 74 & -13.77 & 40.36 & A400 &  & \\
XMMU J025801.6+060148 & 67 & -13.84 & 40.29 & A400 &  & \\
XMMU J010955.6-455551 & 1860 & -12.82 & 41.85 & A2877 & IC1633 & 0.024240 \\
XMMU J011050.6-460013 & 1245 & -13.06 & 41.09 & A2877 & GALEX J011050.4-460013.8 & \\
XMMU J010914.4-455914 & 443 & -13.02 & 41.13 & A2877 & 2XMM J010914.4-455914 & \\
XMMU J011028.2-460422 & 401 & -12.99 & 41.08 & A2877 & ESO 243- G 049 & 0.022395 \\
XMMU J011119.3-455554 & 375 & -13.78 & 40.36 & A2877 & ESO 243- G 051 & 0.021855 \\
XMMU J011019.9-455120 & 90 & -13.29 & 40.84 & A2877 & 2MASX J01101993-4551184  & 0.023243 \\
XMMU J010904.3-454627 & 97 & -13.27 & 40.87 & A2877 & ESO 243- G 045 & 0.025881 \\
XMMU J011007.5-455554 & 258 & -13.07 & 41.05 & A2877 & 2MASX J01100662-4555544 & 0.024360 \\
XMMU J010942.5-455357 & 49 & -13.90 & 40.24 & A2877 &  & \\
XMMU J011017.7-460404 & 108 & -13.32 & 40.81 & A2877 &   & \\
XMMU J010949.1-460235 & 125 & -13.72 & 40.42 & A2877 &   & \\
XMMU J010952.4-460536 & 130 & -13.47 & 40.67 & A2877 & & \\
XMMU J010916.4-454830 & 65 & -13.58 & 40.56 & A2877 &   & \\
XMMU J011032.0-455337 & 178 & -13.47 & 40.67 & A2877 &   & \\
XMMU J011024.5-454426 & 102 & -13.44 & 40.70 & A2877 &   & \\
XMMU J010849.6-455622 & 91 & -13.86 & 40.29 & A2877 &   & \\
XMMU J010930.7-460307 & 66 & -13.65 & 40.47 & A2877 &  & \\
XMMU J010947.3-455327 & 98 & -13.68 & 40.46 & A2877 &   & \\
XMMU J011026.6-455246 & 72 & -13.60 & 40.54 & A2877 &  & \\
XMMU J010933.6-460400 & 42 & -13.84 & 40.30 & A2877 &   & \\
XMMU J010935.8-460621 & 50 & -13.46 & 40.68 & A2877 &   & \\
XMMU J011022.6-460105 & 86 & -13.67 & 40.47 & A2877 &   & \\
XMMU J010853.8-455850 & 54 & -13.77 & 40.37 & A2877 &   & \\
XMMU J011034.9-455149 & 100 & -13.59 & 40.55 & A2877 &   & \\
XMMU J011008.8-455630 & 149 & -13.37 & 40.76 & A2877 &   & \\
XMMU J011058.2-455318 & 70 & -13.68 & 40.46 & A2877 & & \\
XMMU J011001.9-454917 & 78 & -13.62 & 40.52 & A2877 &   & \\
XMMU J011043.2-460419 & 103 & -13.58 & 40.56 & A2877 &   & \\
XMMU J010937.2-454323 & 42 & -13.61 & 40.53 & A2877 &   & \\
XMMU J010858.9-455136 & 104 & -13.20 & 40.93 & A2877 &   & \\
XMMU J011001.7-460818 & 49 & -13.75 & 40.39 & A2877 &   & \\
XMMU J011001.9-460335 & 63 & -13.97 & 40.17 & A2877 &   & \\
XMMU J011006.8-455234 & 40 & -13.53 & 40.61 & A2877 &   & \\
XMMU J011038.1-455829 & 41 & -13.51 & 40.62 & A2877 &   & \\
XMMU J010916.1-455149 & 44 & -13.51 & 40.63 & A2877 &   & \\

\hline
\vspace{0,5mm}
\label{t7}
\end{tabular}
}
\end{table}

\clearpage

\begin{table}
\caption{Properties of point sources studied in this work.}
\vspace{3mm}
\centering
\scalebox{0.92}{
\begin{tabular}{ccccccc}
\hline
\hline
Object Name & Net Counts & Log Flux & Log Luminosity & Cluster & Optical Counterpart & Redshift  \\
 & & erg cm$^{-2}$ s$^{-1}$ & erg s$^{-1}$ & & & \\
 \hline
 \hline
XMMU J114505.0+193622 & 3182 & -12.28 & 41.76 & A1367 & NGC 3862 & 0.021718  \\
XMMU J114435.0+195131 & 628 & -13.07 & 40.97 & A1367 &   & \\
XMMU J114409.4+195009 & 576 & -13.24 & 40.81 & A1367 &   & \\
XMMU J114448.9+194742 & 454 & -12.86 & 41.17 & A1367 &  NGC 3860 & 0.018663 \\
XMMU J114427.1+194338 & 303 & -13.32 & 40.72 & A1367 &   & \\
XMMU J114452.5+195133 & 515 & -13.02 & 41.02 & A1367 &   & \\
XMMU J114515.8+194951 & 136 & -13.38 & 40.66 & A1367 &   & \\
XMMU J114435.5+195029 & 146 & -13.52 & 40.52 & A1367 &   & \\
XMMU J114503.8+195826 & 103 & -13.48 & 40.56 & A1367 & NGC 3861 & 0.01690 \\
XMMU J114436.5+195336 & 187 & -13.48 & 40.56 & A1367 &   & \\
XMMU J114454.7+194634 & 175 & -13.30 & 40.74 & A1367 & CGCG 097-125 & 0.027436  \\
XMMU J114457.6+195302 & 145 & -13.59 & 40.45 & A1367 &   & \\
XMMU J114436.5+193831 & 117 & -13.52 & 40.52 & A1367 &   & \\
XMMU J114410.7+195327 & 100 & -13.57 & 40.47 & A1367 &   & \\
XMMU J114359.2+193955 & 46 & -13.42 & 40.61 & A1367 &   & \\
XMMU J114439.1+194525 & 167 & -13.43 & 40.61 & A1367 &   & \\
XMMU J114416.8+194417 & 44 & -13.48 & 40.56 & A1367 &   & \\
XMMU J114420.1+195849 & 41 & -13.56 & 40.48 & A1367 &   & \\
XMMU J114438.1+194405 & 153 & -13.45 & 40.59 & A1367 &   & \\
XMMU J114459.7+194742 & 96 & -13.62 & 40.41 & A1367 &   & \\
XMMU J114422.1+193937 & 76 & -13.46 & 40.58 & A1367 &   & \\
XMMU J114526.6+194345 & 43 & -13.55 & 40.49 & A1367 &   & \\
XMMU J114420.9+195508 & 53 & -13.71 & 40.32 & A1367 &   & \\
XMMU J114507.7+193552 & 74 & -13.49 & 40.55 & A1367 &   & \\
XMMU J114447.3+194621 & 80 & -13.69 & 40.35 & A1367 & NGC 3860B & 0.02825 \\
XMMU J114501.7+194549 & 93 & -13.91 & 40.13 & A1367 &   & \\
XMMU J114536.9+195304 & 42 & -13.84 & 40.20 & A1367 &   & \\
XMMU J114537.7+195330 & 54 & -13.57 & 40.46 & A1367 &   & \\
XMMU J114507.7+195419 & 40 & -13.54 & 40.50 & A1367 & & \\
XMMU J114507.7+195757 & 39 & -13.37 & 40.66 & A1367 &  & \\
XMMU J114508.4+194905 & 38 & -13.57 & 40.46 & A1367 & & \\ 
XMMU J114402.2+195700 & 153 & -13.50 & 40.55 & A1367 & NGC 3842 & 0.021068 \\
XMMU J114359.3+195632 & 59 & -13.48 & 40.56 & A1367 & GALEXASC J114359.29+195633.6 & 0.023323 \\
XMMU J140729.8-270104 & 128 & -12.46 & 41.58  & A3581 & IC 4374 & 0.021798 \\
XMMU J140714.2-270027 & 289 & -13.63 & 40.45 & A3581 & & \\
XMMU J140715.6-270932 & 1947 & -12.89 & 41.24 & A3581 & ESO 510- G 066 & 0.024333 \\
XMMU J140827.1-265828 & 106 & -13.36 & 40.72 & A3581 & & \\
XMMU J140656.6-265158 & 86 & -13.79 & 40.29 & A3581 & & \\
XMMU J140819.2-270150 & 109 & -13.55 & 40.52 & A3581 & & \\
XMMU J140751.8-265827 & 114 & -13.97 & 40.11 & A3581 & & \\
XMMU J140750.9-271138 & 130 & -13.86 & 40.22 & A3581 & & \\
XMMU J140701.2-265554 & 104 & -13.87 & 40.21 & A3581 & & \\
XMMU J140758.8-270425 & 91 & -13.84 & 40.24 & A3581 & & \\
XMMU J140751.4-271317 & 73 & -13.63 & 40.45 & A3581 & & \\
XMMU J140654.2-265314 & 56 & -13.93 & 40.15 & A3581 & & \\
XMMU J140737.0-270700 & 52 & -13.99 & 40.09 & A3581 & & \\
XMMU J140646.1-270112 & 69 & -13.82 & 40.25 & A3581 & & \\
XMMU J140743.0-265838 & 168 & -14.00 & 40.08 & A3581 & & \\
XMMU J140649.2-270031 & 51 & -13.98 & 40.09 & A3581 & & \\
XMMU J140720.9-265151 & 73 & -13.58 & 40.49 & A3581 & & \\
XMMU J140754.2-270555 & 118 & -13.98 & 40.09  & A3581 & & \\
XMMU J140805.5-270617 & 83 & -13.78 & 40.30  & A3581 & & \\
XMMU J140721.6-264716 & 134 & -12.77 & 41.30 & A3581 & & \\
XMMU J140720.6-270709 & 103 & -13.68 & 40.39 & A3581 & & \\
XMMU J140659.5-265213 & 58 & -13.70 & 40.37 & A3581 & & \\
XMMU J140803.6-270841 & 54 & -13.49 & 40.58 & A3581 & & \\
XMMU J140721.6-265328 & 50 & -13.78 & 40.29  & A3581 & & \\
XMMU J140643.9-265710 & 51 & -13.87 & 40.20 & A3581 & & \\
XMMU J140705.8-271047 & 71 & -13.92 & 40.16 & A3581 & & \\
XMMU J140811.5-265911 & 49 & -13.83 & 40.24  & A3581 & & \\
XMMU J140718.7-265415 & 97 & -13.66 & 40.42 & A3581 & & \\
XMMU J140825.4-270849 & 85 & -13.25 & 40.82 & A3581 & & \\
XMMU J140743.9-270648 & 102 & -13.78 & 40.29 & A3581 & & \\
XMMU J140759.8-271442 & 50 & -13.29 & 40.78 & A3581 & & \\
XMMU J140818.7-270533 &46  & -13.75 & 40.33 & A3581 & & \\
XMMU J140801.4-271340 & 54 & -13.68 & 40.39 & A3581 & & \\
XMMU J140712.0-265007 & 51 & -13.94 & 40.24 & A3581 & ESO 510- G 065 & 0.025671 \\
\hline
\vspace{0,5mm}
\label{t8}
\end{tabular}
}
\end{table}

\clearpage

\begin{table}
\caption{Properties of point sources studied in this work.}
\vspace{3mm}
\centering
\scalebox{0.92}{
\begin{tabular}{ccccccc}
\hline
\hline
Object Name & Net Counts & Log Flux & Log Luminosity & Cluster & Optical Counterpart & Redshift  \\
 & & erg cm$^{-2}$ s$^{-1}$ & erg s$^{-1}$ & & & \\
 \hline
 \hline
XMMU J140840.1-270219 & 41 & -13.74 & 40.33 & A3581 & & \\
XMMU J140800.7-270542 & 55 & -13.95 & 40.12  & A3581 & & \\
XMMU J140844.4-270024 & 38 & -13.62 & 40.46 & A3581 & & \\
XMMU J140652.8-265631 & 186 & -13.19 & 40.88  & A3581 & & \\
XMMU J140806.5-270433 & 95 & -13.67 & 40.41 & A3581 & & \\
XMMU J140809.4-265902 & 81 & -13.64 & 40.44 & A3581 & & \\
XMMU J140752.1-264804 & 95 & -13.08 & 40.99 & A3581 & & \\
XMMU J140757.1-265318 & 36 & -13.40 & 40.67 & A3581 & & \\
XMMU J140741.5-265121 & 54 & -13.41 & 40.66 & A3581 & & \\
XMMU J140650.2-270206 & 49 & -13.94 & 40.14 & A3581 & & \\
XMMU J140645.6-265433 & 39 & -13.63 & 40.44 & A3581 & & \\
XMMU J140750.9-270622 & 49 & -13.92 & 40.15 & A3581 & & \\
XMMU J140632.6-270045 & 43 & -13.88 & 40.20 & A3581 & & \\
XMMU J152305.3+083631 & 1237 & -12.71 & 41.72 & A2063 & CGCG 077-097 & 0.034174 \\
XMMU J152252.6+083735 &102 & -12.95 & 41.49 & A2063 & & \\
XMMU J152318.7+084319 & 67 &  -13.73 & 40.72 & A2063 & & \\
XMMU J152342.5+084535 & 64 & -13.22 & 41.23 & A2063 & & \\
XMMU J152249.9+083643 & 201 & -13.28 & 41.16 & A2063 & & \\
XMMU J152323.5+083212 & 107 & -13.47 & 40.97 & A2063 & & \\
XMMU J152331.2+082744 & 42 & -13.86 & 40.59 & A2063 & & \\
XMMU J152326.9+083428 & 265 & -13.34 & 41.11 & A2063 & & \\
XMMU J152250.9+084447 & 34 & -13.21 & 41.21 & A2063 & & \\
XMMU J152256.6+083858 & 38 & -13.59 & 40.86 & A2063 & 2MASX J15225650+0839004 & 0.03361 \\
XMMU J152237.4+083530 & 42 & -13.59 & 40.85 & A2063 & & \\
XMMU J152248.0+082759 & 87 & -13.61 & 40.84 & A2063 & & \\
XMMU J152226.9+083556 & 35 & -13.93 & 40.52 & A2063 & & \\
XMMU J152322.3+082159 & 46 & -13.22 & 41.23 & A2063 & & \\
XMMU J152248.0+084306 & 44 & -13.94 & 40.51 & A2063 & & \\
XMMU J152359.5+084117 & 32 & -13.55 & 40.90 & A2063 & & \\
XMMU J152404.6+084115 & 31 & -13.59 & 40.86 & A2063 & & \\
XMMU J152312.5+083259 & 1820 & -12.10 & 42.40 & A2063 & 2MASX J15231224+0832590 & 0.036619 \\
XMMU J152327.1+083553 & 53 & -13.22 & 41.27 & A2063 & & \\
XMMU J152240.6+082621 & 47 & -13.28 & 41.20 & A2063 & & \\
XMMU J152234.3+082854 & 31 & -13.94 & 40.55 & A2063 & & \\
XMMU J152225.9+084134  & 30 & -13.92 & 40.57 & A2063 & & \\
XMMU J152219.7+083803 & 94 & -13.59 & 40.90 & A2063 & & \\
XMMU J152224.7+084244 & 33 & -13.66 & 40.82 & A2063 & & \\
XMMU J113449.4+490438 & 357 & -13.73 & 40.68 & A1314 & IC  712 & 0.03335 \\
XMMU J113450.2+490326 & 344 & -13.76 & 40.65 & A1314 &  & \\
XMMU J113439.1+490623 & 53 & -13.95 & 40.46 & A1314 &  & \\
XMMU J113409.1+491516 & 48 & -12.22 & 42.28 & A1314 & 2MASX J11340896+4915162 & 0.03723 \\
XMMU J113543.9+490215 & 273 & -12.13 & 42.15  & A1314 & MCG+08-21-065 & 0.02967 \\
XMMU J113421.6+490050 & 334 & -13.47 & 40.95 & A1314 &  & \\
XMMU J113446.6+485721 & 200 & -13.64 & 40.72 & A1314 & IC  711 & 0.03160 \\
XMMU J113447.3+490133 & 301 & -13.73 & 40.69 & A1314 &  & \\
XMMU J113359.3+490343 & 298 & -13.27 & 41.09 & A1314 & IC  708 & 0.03165 \\
XMMU J113408.4+490318 & 207 & -13.16 & 41.24 & A1314 &  & \\
XMMU J113425.0+490647 & 111 & -13.50 & 40.90 & A1314 &  & \\
XMMU J113359.5+491246 & 69 & -13.76 & 40.65 & A1314 & LEDA 97398 & 0.031600 \\
XMMU J113509.1+490658 & 91 & -13.88 & 40.54 & A1314 &  & \\
XMMU J113441.8+490918 & 118 & -13.86 & 40.56 & A1314 &  & \\
XMMU J113451.4+491203 & 137 & -13.56 & 42.36  & A1314 &  & \\
XMMU J113548.5+491150 & 49 & -13.48 & 40.93 & A1314 &  & \\
XMMU J113355.9+490955 & 61 & -13.32 & 41.08 & A1314 &  & \\
XMMU J113424.7+491230 & 38 &  -13.85 & 40.55 & A1314 &  & \\
XMMU J113425.9+490909 & 81 & -13.83 & 40.59 & A1314 &  & \\
XMMU J113431.4+485354 & 34 & -13.91 & 40.50 & A1314 &  & \\
XMMU J113454.2+485506 & 30 & -13.87 & 40.53 & A1314 &  & \\
XMMU J113404.6+485519 & 76 & -13.38 & 41.02 & A1314 &  & \\
XMMU J113331.2+485920 & 66 & -13.85 & 40.57 & A1314 &  & \\
XMMU J113412.2+485614 & 38 & -13.99 & 40.41 & A1314 &  & \\
XMMU J113420.9+490937 & 47 & -13.89 & 40.52 & A1314 &  & \\
XMMU J113424.7+491623 & 31 & -13.76 & 40.65 & A1314 &  & \\
XMMU J113526.2+485555 & 30 & -13.77 & 40.62 & A1314 &  & \\
XMMU J113533.8+485620 & 34 & -13.94 & 40.47 & A1314 &  & \\
XMMU J113317.3+490607 & 79 & -13.19 & 41.22 & A1314 &  & \\

\hline
\vspace{0,5mm}
\label{t9}
\end{tabular}
}
\end{table}

\clearpage

\begin{table}
\caption{Properties of point sources studied in this work.}
\vspace{3mm}
\centering
\scalebox{0.92}{
\begin{tabular}{ccccccc}
\hline
\hline
Object Name & Net Counts & Log Flux & Log Luminosity & Cluster & Optical Counterpart & Redshift  \\
 & & erg cm$^{-2}$ s$^{-1}$ & erg s$^{-1}$ & & & \\
 \hline
 \hline
XMMU J113328.3+491218 & 48 & -13.66 & 40.75 & A1314 &  & \\
XMMU J113517.8+491259 & 40 & -13.82 & 40.59 & A1314 &  & \\
XMMU J113334.6+485659 & 35 & -13.86 & 40.55 & A1314 &  & \\
XMMU J113347.3+490614 & 37 & -13.96 & 40.45 & A1314 &  & \\
XMMU J113330.0+490901 & 31 & -13.77 & 40.63 & A1314 &  & \\
XMMU J113423.0+490326 & 42 & -13.43 & 40.97 & A1314 &  & \\
XMMU J113408.6+485932 & 39 & -13.01 & 41.39  & A1314 &  & \\
XMMU J113406.5+490622 & 30 & -13.55 & 40.85 & A1314 &  & \\
XMMU J113539.8+485459 & 143 & -13.05 & 41.36 & A1314 &  & \\ 
XMMU J113554.7+485809 & 31 & -13.16 & 41.26 & A1314 &  & \\
XMMU J140141.8-113625 & 589 & -13.20 & 41.31 &  A1836 & MCG -02-36-002 & 0.037776 \\
XMMU J140139.4-113749 & 197 & -13.37 & 41.11 & A1836 & & \\
XMMU J140147.8-113438 & 31 & -13.65 &  40.81 & A1836 & & \\
XMMU J140131.4-113309 & 37 & -13.81 & 40.66 & A1836 & & \\
XMMU J140215.6-113748 & 417 & -12.63 & 41.84 & A1836 & & \\
XMMU J140207.9-113553 & 411 & -13.42 & 41.06 & A1836 & & \\
XMMU J140210.1-114131 & 164 & -13.90 & 40.59 & A1836 & & \\
XMMU J140143.0-113059 & 85 & -13.23 & 41.23 & A1836 & & \\
XMMU J140207.2-114351 & 62 & -13.75 & 40.73 & A1836 & & \\
XMMU J140157.8-113927 & 71 & -13.21 & 41.06 & A1836 & & \\
XMMU J140145.4-113133 & 66 & -13.64 & 40.84 & A1836 & & \\
XMMU J140123.0-114535 & 29 & -13.94 & 40.54 & A1836 & & \\
XMMU J140131.9-113925 & 31 & -13.46 & 41.12 & A1836 & 2MASX J14013206-1139261 & 0.041662 \\
XMMU J140124.5-112832 & 30 & -13.81 & 40.67 & A1836 & & \\
XMMU J140200.2-112752 & 94 & -13.64 & 40.84 & A1836 & & \\
XMMU J140142.5-112754 & 84 & -13.63 & 40.86 & A1836 & & \\
XMMU J140155.9-112952 & 101 & -13.81 & 40.67 & A1836 & & \\
XMMU J140135.5-112708 & 32 & -13.39 & 41.05 & A1836 & & \\
XMMU J140212.5-114443 & 33 & -13.58 & 40.88 & A1836 & & \\
XMMU J140125.4-114717 & 39 & -13.51 & 40.96 & A1836 & & \\
XMMU J140057.7-113746 & 32 & -13.46 & 41.00 & A1836 & & \\
XMMU J140102.0-113316 & 69 & -13.46 & 41.12 & A1836 & & \\
XMMU J140107.7-112915 & 34 & -13.44 & 41.14 & A1836 & & \\
XMMU J140042.4-113447 & 39 & -13.03 & 41.56 & A1836 & & \\
XMMU J140043.8-113731 & 38 & -13.42 & 41.18 & A1836 & & \\
XMMU J140149.0-113837 & 501 & -12.91 & 41.70 & A1836 & & \\
XMMU J140201.7-113408 & 51 & -13.56 & 41.03 & A1836 & & \\
XMMU J140155.7-113808 & 54 & -13.91 & 40.59 & A1836 & 2MASX J14015570-1138043 & 0.036979 \\
XMMU J140057.6-114051 & 49 & -13.68 & 40.82 & A1836 & & \\
XMMU J140230.7-113646 & 42 & -13.80 & 40.70 & A1836 & & \\
XMMU J140217.3-112612 & 29 & -13.91 & 40.59 & A1836 & & \\
XMMU J140130.5-112247 & 37 & -13.72 & 40.78 & A1836 & & \\
XMMU J011146.5-613139 & 445 & -13.92 & 40.29 & AS137 & NGC 0432 & 0.026929 \\
XMMU J011142.8-613106 & 448 & -13.64 & 40.56 & AS137 & & \\
XMMU J011205.4-613255 & 328 & -13.47 & 40.73 & AS137 & & \\
XMMU J011208.8-613214 & 238 & -13.68 & 40.52 & AS137 & & \\
XMMU J011150.0-613348 & 103 & -13.89 & 40.30 & AS137 & & \\
XMMU J011134.8-613414 & 88 & -13.92 & 40.27 & AS137 & & \\
XMMU J011142.1-612649 & 641 & -13.19 & 41.01 & AS137 & & \\
XMMU J011211.4-612126 & 188 & -13.71 & 40.49 & AS137 & & \\
XMMU J011014.3-613852 & 177 & -12.91 & 41.29 & AS137 & & \\
XMMU J011120.3-612958 & 159 & -13.70 & 40.50 & AS137 & & \\
XMMU J011222.9-612811 & 104 & -13.81 & 40.39 & AS137 & & \\
XMMU J011116.1-612822 & 122 & -13.98 & 40.21 & AS137 & & \\
XMMU J011123.1-612054 & 83 & -13.34 & 40.85 & AS137 & & \\
XMMU J011127.4-612612 & 104 & -13.49 & 40.70 & AS137 & & \\
XMMU J011210.5-612807 & 110 & -13.99 & 40.21 & AS137 & & \\
XMMU J011207.4-614033 & 99 & -13.95 & 40.25 & AS137 & & \\
XMMU J011109.0-613049 & 81 & -13.49 & 40.70 & AS137 & & \\
XMMU J011115.6-613242 & 111 & -13.65 & 40.55 & AS137 & & \\
XMMU J011211.3-612439 & 92 & -13.53 & 40.66 & AS137 & & \\
XMMU J011128.8-613519 & 94 & -13.71 & 40.48 & AS137 & & \\
XMMU J011154.6-614035 & 62 & -13.73 & 40.46  & AS137 & & \\
XMMU J011136.7-612338 & 79 & -13.90 & 40.29 & AS137 & & \\
XMMU J011144.0-614135 & 38 & -13.79 & 40.40 & AS137 & & \\
XMMU J011325.0-612746 & 67 & -13.99 & 40.21 & AS137 & & \\
XMMU J011201.8-612157 & 55 & -13.89 & 40.30 & AS137 & & \\
\hline
\vspace{0,5mm}
\label{t10}
\end{tabular}
}
\end{table}

\clearpage

\begin{table}
\caption{Properties of point sources studied in this work.}
\vspace{3mm}
\centering
\scalebox{0.92}{
\begin{tabular}{ccccccc}
\hline
\hline
Object Name & Net Counts & Log Flux & Log Luminosity & Cluster & Optical Counterpart & Redshift  \\
 & & erg cm$^{-2}$ s$^{-1}$ & erg s$^{-1}$ & & & \\
 \hline
 \hline
XMMU J011105.5-612548 & 141 & -13.02 & 41.16 & AS137 & & \\
XMMU J010949.4-613153 & 174 & -13.06 & 41.13 & AS137 & & \\
XMMU J011220.2-613752 & 83 & -13.57 & 40.62 & AS137 & & \\
XMMU J011059.5-613832 & 52 & -13.94 & 40.25 & AS137 & & \\
XMMU J011114.5-612922 & 48 & -13.78 & 40.41 & AS137 & & \\
XMMU J011251.4-613959 & 65 & -13.77 & 40.43 & AS137 & 2MASX J01125179-6139513 & 0.026442 \\
XMMU J011058.3-612752 & 107 & -13.48 & 40.71 & AS137 & & \\
XMMU J011318.7-613133 & 64 & -13.33 & 40.86 & AS137 & & \\
XMMU J011022.1-613043 & 46 & -13.70 & 40.49 & AS137 & & \\
XMMU J011234.0-613615 & 29 & -13.73 & 40.46 & AS137 & & \\
XMMU J011054.8-612940 & 42 & -13.83 & 40.36 & AS137 & & \\
XMMU J011315.5-613609 & 47 & -13.67 & 40.52 & AS137 & & \\
XMMU J011029.9-612637 & 37 & -13.92 & 40.27 & AS137 & & \\
XMMU J011244.1-612341 & 30 & -13.71 & 40.48 & AS137 & & \\
XMMU J011255.6-613655 & 31 & -13.62 & 40.57 & AS137 & & \\
XMMU J011249.7-612938 & 39 & -13.89 & 40.30 & AS137 & & \\
XMMU J011340.9-613101 & 38 & -13.49 & 40.70 & AS137 & & \\
XMMU J011247.9-613645 & 35 & -13.95 & 40.25 & AS137 & & \\
XMMU J011157.1-613920 & 31 & -13.89 & 40.30 & AS137 & & \\
XMMU J011048.5-614303 & 30 & -13.72 & 40.48 & AS137 & & \\
XMMU J011202.3-613006 & 93 & -13.47 & 40.72 & AS137 & & \\
XMMU J011142.8-612813 & 85 & -13.69 & 40.51 & AS137 & & \\
XMMU J011213.1-612015 & 81 & -13.36 & 40.84 & AS137 & & \\
XMMU J011314.2-612331 & 65 & -13.41 & 40.78 & AS137 & & \\
XMMU J011231.9-611906 & 37 & -13.55 & 40.64 & AS137 & & \\
XMMU J011220.4-613648 & 50 & -13.65 & 40.55 & AS137 & & \\
XMMU J011314.3-612249 & 37 & -13.46 & 40.73 & AS137 & & \\
XMMU J010959.0-612601 & 28 & -13.46 & 40.73 & AS137 & & \\
XMMU J011007.3-612417 & 30 & -13.71 & 40.49 & AS137 & & \\
XMMU J141220.9-342022 & 1460 & -13.83 & 40.72 & AS758 & MCG-06-31-029 & 0.03850 \\
XMMU J141117.5-341116 & 302 & -13.09 & 41.42 & AS758 & & \\
XMMU J141216.6-342422 & 988 & -13.06 & 41.46 & AS758 & & \\
XMMU J141226.2-341715 & 746 & -13.49 & 41.03 & AS758 & & \\
XMMU J141201.2-341548 & 430 & -13.50 & 41.03 & AS758 & & \\
XMMU J141308.6-342105 & 197 & -13.01 & 41.49 & AS758 & & \\
XMMU J141200.2-341657 & 272 & -13.85 & 40.68 & AS758 & & \\
XMMU J141157.6-341840 & 303 & -13.92 & 40.61 & AS758 & & \\
XMMU J141231.7-342141 & 363 & -13.79 & 40.74 & AS758 & & \\
XMMU J141124.7-342353 & 248 & -13.59 & 40.94 & AS758 & & \\
XMMU J141253.8-342638 & 243 & -13.85 & 40.68 & AS758 & & \\
XMMU J141312.2-342051 & 123 & -13.69 & 40.83 & AS758 & & \\
XMMU J141200.0-342245 & 115 & -13.91 & 40.61 & AS758 & & \\
XMMU J141216.8-342556 & 112 & -13.98 & 40.54 & AS758 & & \\
XMMU J141246.8-342532 & 139 & -13.80 & 40.72 & AS758 & & \\
XMMU J141145.6-342441 & 177 & -13.89 & 40.64 & AS758 & & \\
XMMU J141200.7-341036 & 123 & -13.67 & 40.85 & AS758 & & \\
XMMU J141311.3-342216 & 281 & -13.69 & 40.84 & AS758 & & \\
XMMU J141229.3-341741 & 149 & -13.96 & 40.67 & AS758 & 2MASX J14122917-3417417 & 0.04300 \\
XMMU J141134.3-342353 & 71 & -13.89 & 40.63 & AS758 & & \\
XMMU J141210.3-343142 & 64 & -13.31 & 41.20 & AS758 & & \\
XMMU J141259.5-341734 & 93 & -13.79 & 40.72 & AS758 & & \\
XMMU J141241.3-342659 & 135 & -13.80 & 40.72 & AS758 & & \\
XMMU J141149.9-343008 & 102 & -13.90 & 40.63 & AS758 & & \\
XMMU J141127.8-342255 & 66 & -13.81 & 40.71 & AS758 & & \\
XMMU J141246.3-342816 & 63 & -13.71 & 40.80 & AS758 & & \\
XMMU J141314.2-341952 & 88 & -13.62 & 40.89 & AS758 & & \\
XMMU J141215.8-342717 & 30 & -13.74 & 40.77 & AS758 & & \\
XMMU J141249.7-342552 & 62 & -13.82 & 40.69 & AS758 & & \\
XMMU J141120.2-341449 & 63 & -13.55 & 40.96 & AS758 & & \\
XMMU J141207.7-340830 & 67 & -13.88 & 40.64 & AS758 & & \\
XMMU J141232.8-341653 & 70 & -13.97 & 40.54 & AS758 & & \\
XMMU J141118.2-342059 & 35 & -13.68 & 40.82 & AS758 & & \\
XMMU J141230.7-340947 & 757 & -13.56 & 40.98 & AS758 & & \\
XMMU J141255.7-341247 & 362 & -13.78 & 40.75 & AS758 & & \\
XMMU J141216.6-341034 & 179 & -13.50 & 41.01 & AS758 & & \\
XMMU J141301.7-341331 & 63 & -13.85 & 40.67 & AS758 & & \\
XMMU J141241.6-341103 & 119 & -13.90 & 40.62 & AS758 & & \\
\hline
\vspace{0,5mm}
\label{t11}
\end{tabular}
}
\end{table}

\clearpage

\begin{table}
\caption{Properties of point sources studied in this work.}
\vspace{3mm}
\centering
\scalebox{0.92}{
\begin{tabular}{ccccccc}
\hline
\hline
Object Name & Net Counts & Log Flux & Log Luminosity & Cluster & Optical Counterpart & Redshift  \\
 & & erg cm$^{-2}$ s$^{-1}$ & erg s$^{-1}$ & & & \\
 \hline
 \hline

XMMU J141146.6-341906 & 87 & -13.86 & 40.65 & AS758 & & \\
XMMU J141219.9-342557 & 89 & -13.86 & 40.65 & AS758 & & \\
XMMU J141144.2-341353 & 74 & -13.88 & 40.63 & AS758 & & \\
XMMU J141209.8-341327 & 55 & -13.71 & 40.80 & AS758 & & \\
XMMU J141139.1-340615 & 79 & -13.50 & 41.02 & AS758 & & \\
XMMU J141117.0-342710 & 74 & -13.68 & 40.84 & AS758 & & \\
XMMU J141204.1-340538 & 42 & -13.37 & 41.13 & AS758 & & \\
XMMU J141221.6-342716 & 52 & -13.96 & 40.56 & AS758 & & \\
XMMU J141139.6-340810 & 72 & -13.77 & 40.75 & AS758 & & \\
XMMU J141141.0-340707 & 39 & -13.77 & 40.75  & AS758 & & \\
XMMU J141137.0-340736 & 38 & -13.36 & 41.15 & AS758 & & \\
XMMU J141236.7-342930 & 54 & -13.95 & 40.57 & AS758 & & \\
XMMU J141302.4-342438 & 37 & -13.86 & 40.64 & AS758 & & \\
XMMU J141311.8-341709 & 40 & -13.75 & 40.76 & AS758 & & \\
XMMU J141234.3-341849 & 151 & -13.58 & 40.93 & AS758 & & \\
XMMU J141240.1-341617 & 56 & -13.64 & 40.88 & AS758 & & \\
XMMU J141245.4-342343 & 343 & -13.35 & 41.18 & AS758 & & \\
XMMU J141304.3-342557 & 43 & -13.44 & 41.07 & AS758 & & \\
XMMU J141215.6-341628 & 42 & -13.98 & 40.54 & AS758 & & \\
XMMU J141149.9-341643 & 46 & -13.95 & 40.57 & AS758 & & \\
XMMU J141300.2-342947 & 29 & -13.91 & 40.61 & AS758 & & \\
XMMU J141223.0-341330 & 218 & -13.47 & 41.06 & AS758 & & \\
XMMU J231638.6-022527 & 522 & -12.54 & 41.67 & RXCJ2315.7-0222 & & \\
XMMU J231544.4-022254 & 1768 & -13.61 & 40.56 & RXCJ2315.7-0222 & NGC 7556 & 0.025041 \\
XMMU J231624.5-021457 & 310 & -13.35 & 40.80 & RXCJ2315.7-0222 & & \\
XMMU J231553.8-022615 & 235 & -13.78 & 40.38 & RXCJ2315.7-0222 & & \\
XMMU J231623.8-022550 & 132 & -13.54 & 40.61 & RXCJ2315.7-0222 & & \\
XMMU J231523.5-021660 & 130 & -13.67 & 40.48 & RXCJ2315.7-0222 & & \\
XMMU J231637.4-021950 & 110 & -13.83 & 40.32 & RXCJ2315.7-0222 & NGC 7566 & 0.026548 \\
XMMU J231535.8-021727 & 71 & -13.99 & 40.16 & RXCJ2315.7-0222 & & \\
XMMU J231555.9-021644 & 85 & -13.33 & 40.81 & RXCJ2315.7-0222 & & \\
XMMU J231610.3-021128 & 74 & -13.80 & 40.35 & RXCJ2315.7-0222 & & \\
XMMU J231605.0-022533 & 59 & -13.95 & 40.20 & RXCJ2315.7-0222 & & \\
XMMU J231527.6-021525 & 60 & -13.56 & 40.58 & RXCJ2315.7-0222 & & \\
XMMU J231618.2-021330 & 37 & -13.58 & 40.57 & RXCJ2315.7-0222 & & \\
XMMU J231610.3-021502 & 29 & -13.47 & 40.67 & RXCJ2315.7-0222 & & \\
XMMU J231610.3-021238 & 36 & -13.76 & 40.39 & RXCJ2315.7-0222 & & \\
XMMU J231600.0-022943 & 159 & -13.23 & 40.91 & RXCJ2315.7-0222 & & \\
XMMU J231619.7-021351 & 50 & -13.62 & 40.53 & RXCJ2315.7-0222 & & \\
XMMU J231552.3-023008 & 73 & -13.76 & 40.39 & RXCJ2315.7-0222 & & \\
XMMU J231547.8-022841 & 86 & -13.94 & 40.21 & RXCJ2315.7-0222 & & \\
XMMU J231631.4-021247 & 471 & -12.88 & 41.33 & RXCJ2315.7-0222 & & \\
XMMU J231642.2-022452 & 114 & -12.87 & 41.26 & RXCJ2315.7-0222 & & \\
XMMU J231504.6-022254 & 59 & -13.62 & 40.53 & RXCJ2315.7-0222 & & \\
XMMU J231559.0-021042 & 31 & -13.69 & 40.46 & RXCJ2315.7-0222 & & \\
XMMU J231614.4-021426 & 45 & -13.97 & 40.18 & RXCJ2315.7-0222 & & \\
XMMU J231544.6-023059 & 44 & -13.82 & 40.32 & RXCJ2315.7-0222 & & \\
XMMU J231538.9-021608 & 37 & -13.60 & 40.54 & RXCJ2315.7-0222 & & \\
XMMU J231633.8-023017 & 35 & -13.67 & 40.48 & RXCJ2315.7-0222 & & \\
XMMU J231500.7-022515 & 91 & -13.36 & 40.85 & RXCJ2315.7-0222 & & \\
XMMU J231602.2-022532 & 47 & -13.83 & 40.38 & RXCJ2315.7-0222 & & \\
XMMU J231513.7-022235 & 31 & -13.74 & 40.46 & RXCJ2315.7-0222 & & \\
XMMU J231527.1-022812 & 38 & -13.65 & 40.55 & RXCJ2315.7-0222 & & \\
XMMU J231610.3-023034 & 29 & -13.58 & 40.62 & RXCJ2315.7-0222 & & \\
XMMU J231547.8-023050 & 45 & -13.95 & 40.26 & RXCJ2315.7-0222 & & \\
XMMU J231556.6-022415 & 30 & -13.54 & 40.65 & RXCJ2315.7-0222 & & \\
\hline
\vspace{0,5mm}
\label{t12}
\end{tabular}
}
\end{table}


\bsp	
\label{lastpage}
\end{document}